\documentclass{article}
\usepackage[utf8]{inputenc}
\usepackage[a4paper, total={6in, 8in}]{geometry}
\usepackage{graphicx} 
\usepackage{bm}
\usepackage{amsfonts,amssymb}
\usepackage[dvipsnames]{xcolor}
\usepackage{hyperref}
\hypersetup{colorlinks=true, linkcolor=blue, anchorcolor=blue, citecolor=blue}
\usepackage[round]{natbib}
\usepackage{bbding}
\usepackage{amsmath}
\usepackage{pifont}
\usepackage{bbm}
\usepackage{setspace}
\usepackage{tabularx}
\usepackage{siunitx}
\usepackage{booktabs} 
\usepackage{multirow}
\usepackage{adjustbox}
\usepackage{makecell}
\usepackage{threeparttable}
\usepackage{appendix}
\usepackage{rotating}
\usepackage{graphicx}
\usepackage{subcaption}
\usepackage{float}
\usepackage{indentfirst}
\usepackage[ruled,vlined]{algorithm2e}
\usepackage{algpseudocode}
\SetKwInput{KwInput}{Input}               
\SetKwInput{KwOutput}{Output}            
\SetKwInput{KwProcedure}{Procedure}

\definecolor{DarkGreen}{RGB}{1,50,32}

\usepackage[affil-it]{authblk}
\title{Mortality Forecasting under Climate Risk: A Stochastic Approach with Distributed Lag Non-Linear Models}

\author[1,*]{Jiacheng Min}
\author[1]{Han Li}
\author[2]{Thomas Nagler}
\author[1]{Shuanming Li}

\affil[1]{\small Department of Economics, The University of Melbourne, VIC 3010, Australia}
\affil[2]{\small Department of Statistics, LMU Munich, Munich Center for Machine Learning, Germany}
\affil[*]{Corresponding author: \href{mailto:jiacheng.min@student.unimelb.edu.au}{jiacheng.min@student.unimelb.edu.au}}

\date{}
\begin{document}
\maketitle

\begin{abstract}
Assessing climate-driven mortality risk has become an emerging area of research in recent decades. 
In this paper, we propose a novel approach to explicitly incorporate climate-driven effects into both single- and multi-population stochastic mortality models. 
The new model consists of two components: a stochastic mortality model, and a distributed lag non-linear model (DLNM). The stochastic component captures the non-climate long-term trend, volatility, and seasonal patterns in mortality rates. The DLNM component captures non-linear and lagged effects of climate variables on mortality, as well as the impact of heat waves and cold waves across different age groups. For model calibration, we propose a novel backfitting algorithm that allows us to disentangle the climate-driven mortality risk from the non-climate-driven stochastic mortality risk. 
We illustrate the effectiveness and improved  short-term (1--18 month)  forecasting performance of our model against four alternative models, using data from three European regions: Athens, Lisbon, and Rome.
Furthermore, as an application of the proposed modeling framework, we utilize future UTCI data generated from climate models to provide total mortality forecasts into 2045 across these regions under two Representative Concentration Pathway (RCP) scenarios, taking both stochastic mortality improvement trend and climate risk into account. The projections show a noticeable decrease in winter mortality alongside a rise in summer mortality, driven by a general increase in UTCI over time. Although we expect slightly lower overall mortality in the short term under RCP8.5 compared to RCP2.6, a long-term increase in total mortality is anticipated under the RCP8.5 scenario. \\

\end{abstract}

\clearpage 
\newpage 

\section{Introduction}
Climate-driven mortality risks have emerged as a growing concern in recent years.
The mortality impact of various climate factors\footnote{While our study focuses on short-term fluctuations of weather variables, we use the term ``climate'' throughout this paper to maintain the connection to climate change.}, such as temperature \citep{armstrong2006models}, air pollution \citep{dominici2002air}, relative humidity \citep{armstrong2019role}, and heat waves \citep{gasparrini2011impact} has been extensively studied in environmental and epidemiological research.

These studies demonstrate that the impact of climate on mortality extends beyond a simple linear association, involving complex relationships and delayed effects. 
To capture non-linear effects as well as lagged effects, \cite{gasparrini2010distributed} proposed the Distributed Lag Non-Linear Model (DLNM), which employs a cross-basis function to quantify the association between climate variables and mortality. Although there is a rich literature on DLNMs and they have been widely used to explore the temperature-mortality relationship, relatively little research has focused on how they can be utilized for mortality forecasting, particularly in a multi-population setting. A key limitation of classical DLNMs is that they do not allow for uncertainty in long-term mortality trends across age groups or regions, which is typically incorporated into stochastic mortality models.

There is an extensive body of research on mortality within the fields of actuarial science and demography. In particular,  \cite{lee1992modeling} marked the beginning of the era of stochastic mortality modeling and forecasting. Continuous improvement in human longevity has been observed since the mid-20th century, thanks to medical advances and economic growth. However, the increasing frequency and intensity of natural events due to climate change have posed a growing threat to human health and longevity. Therefore, it is becoming important to factor the potential impact of climate change into mortality projections. 
In fact, several pioneering studies have included a climate dimension into mortality modeling. 
For example, \cite{seklecka2017mortality, seklecka2019mortality} proposed a temperature-related mortality model, which utilizes annual Pearson correlation coefficients between temperature and mortality rates. 
However, their approach fails to capture short-term and non-linear temperature effects. For example, temperatures may only affect mortality beyond a certain threshold. To address this limitation, \cite{li2022joint} introduced a bivariate peaks-over-threshold approach to model monthly death counts and extreme temperatures. It provides insights into how extreme hot and cold temperatures affect mortality but lacks the ability to project future mortality scenarios. 
More recently, \cite{GUIBERT2025} implemented a two-stage approach to extrapolate temperature-related deaths from total mortality and thereby adjust the estimates and forecasts of multi-population stochastic models.  
Temperature-related deaths are estimated in the first stage using a DLNM, and adjusted annual mortality is then modeled using the \cite{li2005coherent} model. Due to the two-stage nature of this approach, complete separation between temperature- and non-temperature-related mortality trends is not guaranteed, and interactions between the two components are ignored. Moreover, future mortality projections are conducted only at the annual level, which does not provide a breakdown by summer and winter months.

Our research builds upon these previous studies, and we develop a new framework for high-frequency total  mortality forecasting that explicitly accounts for climate effects while addressing the limitations of existing methods. In particular, we want to disentangle the climate-driven mortality risk from the non-climate-driven stochastic mortality risk (we refer to as inherent mortality risk\footnote{It is worth noting that our definition of inherent mortality risk in this study may differ from those used in previous weather-mortality studies.}). 
To achieve this, the proposed model consists of two components that are estimated jointly in a single step. The first component is a stochastic mortality model, which captures the non-climate long-term mortality trend, volatility, and seasonality. We choose the \cite{lee1992modeling} model for the single-population setting and the \cite{li2005coherent} model for the multi-population setting. For the second component, we propose a mixed-frequency DLNM to capture and quantify the climate-driven effects on mortality rates, utilizing publicly available data. These two components are integrated and calibrated through a novel backfitting algorithm, inspired by the work by \cite{friedman1981projection}, allowing for an adaptive and iterative estimation process.  
For the DLNM component, we use the Universal Thermal Climate Index (UTCI) as the exposure variable, as it combines air temperature, wind speed, humidity, and radiation into a single measure \citep{fiala2012utci, blazejczyk2013introduction}.
Additionally, we consider consecutive heat and cold waves in the DLNM component to capture non-additive extreme climate effects. 
This joint model framework, which we call the DLNM--Lee--Carter model or the DLNM--Li--Lee model\footnote{Throughout the paper, we refer the two models as DLNM--LC model and DLNM--LL model, respectively.}, allows for a clear separation of climate-driven mortality components from inherent mortality components, providing more accurate  total  mortality forecasts and further insights into mortality improvement patterns. Future mortality can be projected under different climate scenarios, such as Representative Concentration Pathway (RCP) scenarios \citep{gasparrini2017projections} and Shared Socio-economic Pathway (SSP) scenarios \citep{chen2024impact}. Finally, the proposed model enables weekly mortality projections under different RCP scenarios, utilizing future UTCI data generated from climate models. 

We conduct an empirical study based on region-specific daily UTCI and weekly mortality rates in Athens, Lisbon, and Rome during the period 2015--2019. 
We apply both DLNM--LC and DLNM--LL models to this dataset. 
Our results show that the DLNM component effectively captures the seasonal pattern of UTCI-related mortality rates, where we observe a U-shaped relationship between UTCI and relative mortality risk, indicating elevated risks at both extremely low and high index values. 
The stochastic component, on the other hand, captures the long-term trend in mortality improvement over time. 
We examine the performance of the proposed models via in-sample fitting and expanding window cross-validations. Both proposed models achieve the highest forecasting accuracy in most cases against alternative models, demonstrating their advantages in capturing climate effects in mortality modeling and forecasting.
As a demonstration of the proposed framework, we also produce projections of total mortality through 2045 under two extreme RCP scenarios. 
The results illustrate that under RCP8.5, which is the worst-case emission scenario, substantial mortality reductions in winter could offset the increase in summer mortality in the short run, leading to an overall lower mortality level compared to RCP2.6. However, in the long run, we observe  a reversal of these effects based on the projections, resulting in an overall increase in total mortality.

The main motivation of our model are to provide accurate mortality forecasts at a weekly frequency and to conduct \textcolor{black}{total} mortality projections under different climate scenarios, building on existing knowledge of stochastic mortality models and climate-attributable mortality through DLNMs. The contributions of our paper can be summarized as follows:
\vspace{-0.08in}
\begin{itemize}
\item First, we are among the first to introduce a joint modeling approach that simultaneously models and forecasts inherent mortality trends and climate-driven effects on mortality rates, across different age groups, and under both single- and multi-population settings. 
Our models capture the inherent mortality trend through  a stochastic component, and incorporates non-linear climate effects through a DLNM component, all within a one-step approach.
\vspace{-0.08in}
\item Second, we introduce a novel backfitting algorithm to estimate the proposed DLNM--LC and DLNM--LL models, which is the first application of backfitting in stochastic mortality modeling. This iterative procedure effectively isolates the climate-driven mortality component from the inherent mortality component.
\vspace{-0.08in}
\item Third, within the proposed framework, we conduct total mortality projections under different RCP scenarios to simultaneously account for future mortality improvements and volatilities, as well as seasonality and uncertainty arising from the climate-driven mortality component.
\end{itemize}

The remainder of this paper is structured as follows. Section \ref{sec:Methodology} introduces the methodological framework, detailing the integration of stochastic mortality models with the DLNM. Section \ref{sec:Data} provides a description and visualization of the data. Section \ref{sec:Empirical results} evaluates the fitting and forecasting performance of the proposed models. Section \ref{sec:Mortality projection under RCP scenarios} presents mortality projections under different RCP scenarios, analyzing the potential long-term impacts of climate change on mortality rates. Section \ref{sec:Conclusion} concludes the paper and outlines directions for future research. All analyses were carried out using R \citep{R}, and we share the full dataset together with annotated R code through  GitHub repository \url{https://github.com/jiachengmin2025/climate_driven_mortality_risks}.

\section{Methodology}
\label{sec:Methodology}
\subsection{Distributed lag non-linear model} \label{sec:2.1}
In the field of epidemiology, \cite{gasparrini2010distributed} proposed a flexible DLNM approach to analyze the non-linear and lagged climate effects on mortality rates. 
The climate exposures commonly used in the DLNM model include temperatures \citep{gasparrini2010distributed,cheng2014temperature,madaniyazi2022assessing, wen2023new, madaniyazi2024seasonality}, air pollutions \citep{dominici2002air}, humidity 
\citep{armstrong2019role}, and heat and cold wave indicators \citep{gasparrini2011impact, guo2018quantifying}.

The DLNM introduced by \cite{gasparrini2010distributed} uses a quasi-Poisson regression with a log link function to capture temperature effects on daily death counts as follows:
\begin{equation}
\label{original_dlnm}
\log \mathbb{E}[y(t)] = \beta_0 + \sum_{\ell = 0}^L s(X_{t - \ell}, \ell) + \bm{z_{t}^\prime}\bm{\beta} , 
\end{equation}
where the response variable $y(t)$ is  the number of deaths on day $t$,  $\beta_0$ is the intercept, and $X_{t - \ell}$ is the exposure (\textit{e.g.}, temperature) at lag $\ell$. $s(X_{t - \ell}, \ell)$ represents a cross-basis smoothing function that captures the lagged climate exposure effect after $\ell$ days, with various types of basis functions, such as the natural splines or B-splines, being applicable. $\bm{z_{t}}$ is a vector of covariates, with $\bm{\beta}$ being the vector of the corresponding coefficients.  $\bm{z_{t}^\prime}\bm{\beta}$ captures the effect of any confounders, including long-term trends and seasonality.

The DLNM approach has traditionally been applied to daily mortality and climate data for specific geographical areas, such as the City of New York \citep{gasparrini2010distributed}. While climate data is naturally high-frequency and widely available, the same is not true for mortality data. In fact, the majority of research on mortality modeling and forecasting has been conducted using annual mortality rates {\citep[\textit{e.g.},][]{lee1992modeling}}, primarily due to data completeness considerations. For privacy protection purposes, most publicly available data sources provide weekly death counts as the most detailed breakdown. The mismatch between the frequencies of mortality and climate data presents a challenge for researchers when implementing the DLNM. 

To address this issue, we introduce a modified mixed-frequency DLNM where mortality data has a lower frequency of index $t$ (week) and climate data has a higher frequency of index $\tau$ (day). Our proposed mixed-frequency DLNM is as follows:

\begin{equation}
\label{dlnm}
\log \mathbb{E}[y(t)] = \beta_0 + \sum_{\ell = 0}^L s(X_{\tau_t-\ell}, \ell; \bm{\nu},\bm{\eta}) + \bm{z_{t}^\prime}\bm{\beta},
\end{equation}
where we set the reference day for each week to be the last day of that week, which means for $t=N$, the reference day $\tau_t=7\times N$. 
To illustrate how $y(t)$ and $X_{\tau_t-l}$ are linked in the model, we present the following time-lag matrix:  

\[
\begin{array}{c@{}c}
    & 
    \hspace{-1.2em}
    \begin{array}{cccc}
        \text{Lag 0} & \text{Lag 1} & \cdots & \text{Lag $L$}
    \end{array} \\[1ex]
\begin{array}{c}
{t = 1} \\
{t = 2} \\
\vdots \\
{t = N}
\end{array}
&
    \left(
    \begin{array}{cccc}
        7 & 6 & \cdots & 7 - L \\
        14 & 13 & \cdots & 14 - L \\
        \vdots & \vdots & \ddots & \vdots \\
        \tau_N & \tau_N - 1 & \cdots & \tau_N - L
    \end{array}
    \right),\\[7ex]
& {\tau_{t}-\ell}
\end{array}
\]

In this study, we choose the natural cubic splines function as our cross-basis function, with degrees of freedom $\bm{\nu} = (\nu_1, \nu_2)$ and coefficients $\bm \eta \in \mathbb{R}^{\nu_1 \nu_2}$ for both dimensions. 
We follow the common practice in the literature and set $L = 21$ days \citep{gasparrini2010distributed, gasparrini2014modeling, guo2016projecting, guo2018quantifying, song2022evaluation}. 

The estimation of Equation (\ref{dlnm}) is based on the Maximum Likelihood Estimation employed in \cite{gasparrini2010distributed}. We consider the bi-dimensional spline function $s(\cdot)$ as a tensor product, representing an exposure–lag–response surface of relative risk that can be analyzed across different lag times or varying levels of UTCI \citep{gasparrini2014modeling}. The surface of exposure--lag--response is constructed of exposure--response and lag--response functions, both modeled by natural cubic splines with specified degrees of freedom. 

To account for the  uncertainty of $\bm{\eta}$ for the cross-basis function, and $\bm{\beta}$ for other covariates, the bootstrapping method is applied.
Let $\hat{\bm{\zeta}} := (\hat{\bm{\eta}}, \hat{\bm{\beta}})$ denote the estimated parameter vector, which we assume following a multivariate normal distribution. 
We apply parametric bootstrapping for $B$ times to obtain bootstrapped DLNM parameters $\bm{\hat\zeta}^*_b = (\bm{\eta}^*_b, \ \bm{\beta}^*_b)$ such that
\begin{equation}
\label{bootstrap}
    \bm{\hat\zeta}^*_b \sim \mathcal{MVN}(\hat{\bm{\mu}}_{\bm{\zeta}}, \hat{\bm{\Sigma}}_{\bm{\zeta}}), \ b = 1,\dots, B,
\end{equation}
where $\hat{\bm{\mu}}_{\bm{\zeta}}$ is the maximum quasi-likelihood estimator, and $\hat{\bm{\Sigma}}_{\bm{\zeta}}$ is the associated asymptotic covariance; see \citet[Chapter 6]{wood2017generalized}. 

In addition to providing a solution to the mismatched frequency problem, our proposed approach addresses two other limitations of the DLNM model.
First, the DLNM generally does not pool data across age groups. 
In the current literature, DLNMs are most commonly applied to the entire age range or different age groups separately.
While it makes sense to have distinct climate effects for different age groups, the long-term trend in mortality may not be captured accurately due to noise in the data when the sample size is small. Second, the classical DLNM may not be ideal for forecasting total mortality. Even when a time trend is incorporated into the model, it is typically represented by a natural cubic-spline function, which provides only a deterministic point forecast. A stochastic trend, however, allows for the quantification of forecast uncertainty and better reflects the inherent variability in long-term mortality projections.

We propose a new modeling framework that integrates DLNM into stochastic mortality models.
The proposed model has two components, namely a stochastic component and a DLNM component. The stochastic component captures the common time trend in mortality across age groups, which can be modeled using a time series approach to account for the uncertainty in long-term mortality forecasting.
The DLNM, as the second component of the model, is applied to capture mortality patterns and risks associated with climate exposure. It should be noted that this component is estimated separately for each age group, without accounting for mortality commonality across ages, in order to reflect age-specific differences in the impact of climate exposure.
In the following subsections, we introduce the proposed integrated framework in both single-population and multi-population settings.

\subsection{Single-population setting: DLNM--LC model}
\label{sec:Single-population setting: DLNM--LC model}
The Lee--Carter model marked the beginning of the era of stochastic mortality modeling, and to date it remains one of the most influential models used for mortality forecasting \citep{lee1992modeling}. 
The model is formulated as follows:
\begin{equation}
\log m(x,t) = a(x) + b(x) \kappa(t) + \epsilon(x,t),
\end{equation}
for $x = 1, \dots, N$ and $t = 1, \dots T$, where $m(x,t)$ is the mortality rate for age group $x$ at time $t$, $a(x)$ represents the age-specific mean of mortality rate over time. $b(x)\kappa(t)$ represents the mortality improvement over time associated with each age group, where $\kappa(t)$ represents the common time trend of mortality rates across all age groups, and $b(x)$ represents the age-specific loading to $\kappa(t)$. $\epsilon(x,t)$ is the error term of the model. The following constraints are imposed to solve the model identification issue:
\begin{equation}
\sum_{x=1}^N b(x) = 1 \ \text{and} \ \sum_{t=1}^T \kappa(t) = 0.
\end{equation}

The Lee--Carter model is estimated using Singular Value Decomposition (SVD). Details are provided in online supplementary material Section A.
We define the estimated parameter set of the Lee–Carter model as
$\bm{\Theta}_{\text{LC}} := \left\{\hat{\bm{a}}, \hat{\bm{b}}, \hat{\bm{\kappa}}\right\}$.
Note that the original Lee--Carter model does not include exogenous variables, relying purely on age and time factors to explain and forecast mortality rates. As a result, the model cannot directly account for external drivers of mortality change, such as climate factors.

Under the single population setting,  we adopt the Lee--Carter model as the stochastic mortality component, and define the DLNM component for age $x$ as:   
\begin{align}
\mathcal{S}_x\left(U_{\tau_t}, \dots, U_{\tau_t-L}, \text{HWD}_t, \text{CWD}_t \right) &:= \beta_0 + \sum_{\ell = 1}^L s(U_{\tau_t-\ell}, \ell; \bm{\nu}, \bm{\eta}) + \beta_1\text{HWD}_t + \beta_2\text{CWD}_t,
\end{align}
where we choose Universal Thermal Climate Index as our climate exposure variable, denoted by $U$.  
$\text{HWD}_t$ and $\text{CWD}_t$ are the number of heatwave days and cold wave days during week $t$, with corresponding coefficients $\beta_1$ and $\beta_2$, respectively (see detailed definitions in Section \ref{sec:Heat wave and cold wave variable}).

Putting the two components together, the proposed DLNM--LC model is specified as follows:
\begin{equation}
\label{dlnmlc_expression}
\log m(x,t) = \underbrace{a(x) + b(x) \kappa(t)}_{\text{Lee--Carter component}} + \underbrace{\mathcal{S}_x\left(U_{\tau_t}, \dots, U_{\tau_t-L}, \text{HWD}_t, \text{CWD}_t\right)}_{\text{DLNM component}} +  \epsilon(x,t),
\end{equation}
where $a(x)$, $b(x)$, and $\kappa(t)$ represent the Lee–Carter model parameters, referred to as the stochastic mortality component. 
The term $\mathcal{S}_x\left(U_{\tau_t}, \dots, U_{\tau_t-L}, \text{HWD}_t, \text{CWD}_t\right)$, denoted as the DLNM component, characterizes the non-linear relationship between UTCI and mortality rates for age group $x$. $\epsilon(x,t)$ is the error term of the model. No specific distributional assumption is imposed on the errors, other than that they are independent of the covariates, have a mean of zero, and have finite variance.

A related model combining stochastic population mortality dynamics with a temperature-related DLNM component was recently proposed by \citet{GUIBERT2025}. The key differences to our approach concern the temporal resolution (mixed daily/annual versus weekly) and the estimation strategy.\footnote{For a systematic comparison between our approach and the two-stage approach by \cite{GUIBERT2025}, please refer to online supplementary material Section G.2.}
Specifically, \citet{GUIBERT2025} employ a rather ad-hoc two-stage procedure that ignores interactions between the two components. In the first stage, the temperature-related effect is estimated from observed death counts, replacing the stochastic mortality component by a coarse deterministic trend. In the second stage, the residual deaths not explained by temperature are modeled using SVD. This procedure disregards information from the stochastic mortality dynamics when estimating the temperature component, and the inconsistent treatment of population dynamics across stages precludes a coherent decomposition of effects.

To overcome these issues, we propose a new estimation algorithm that simultaneously estimates the two distinct components of the model.
The algorithm is based on the backfitting principle, inspired by \citet{friedman1981projection} for projection pursuit regression, and later extended to generalized additive models by \citet{buja1989linear} and \citet{hardle1993backfitting}. In the context of nonparametric additive models, backfitting is an iterative estimation procedure that repeatedly cycles through each covariate while holding the others fixed, updating its corresponding function until the estimates converge or stop changing significantly. This approach efficiently breaks down a multivariate smoothing problem into a series of univariate ones which makes it computationally feasible, particularly in high-dimensional settings.

For the proposed DLNM--LC model, we propose the following backfitting approach with the goal of jointly estimating the two model components and disentangling climate-driven mortality from non-climate-driven mortality. 
The high level idea can be explained through Equation~\eqref{dlnmlc_expression}. Given the DLNM component, it is straightforward to estimate the Lee--Carter component by applying the SVD algorithm to the log-mortality surface after removing the temperature effect, that is $\log m(x, t) - \text{DLNM}$. Conversely, given the Lee--Carter component, it is straightforward to estimate the DLNM component treating the residual $\log m(x, t) - \text{Lee--Carter component}$ as the response in the UTCI-related model. Iterating between these two estimation steps successively refines both components until the variation explainable by their combined contribution is fully captured.

To make this more precise, denote the original log mortality matrix $\log \bm{m}$ as $\log \hat{\bm{m}}^{(0)}$. The backfitting algorithm is applied as follows. First, we fit a Lee--Carter model on $\log \hat{\bm{m}}^{(0)}$ to obtain the parameter set $\bm{\Theta}^{(0)}_{\text{LC}} = \{\hat{\bm{a}}^{(0)}, \hat{\bm{b}}^{(0)}, \hat{\bm{\kappa}}^{(0)}\}$ in the first recursion, and remove the age-specific mean $\hat{\bm{a}}^{(0)}$ from the log mortality matrix $\log \hat{\bm{m}}^{(0)}$. 
The remaining component is arranged by age group, denoted as partial residuals and defined as $\bm{e}^{(0)} = [\bm{e}^{(0)}_1, \dots, \bm{e}^{(0)}_N] = \log\hat{\bm{m}}^{(0)} - \hat{\bm{a}}^{(0)}$. The partial residuals are then fitted by the DLNM, separately for each age group $x$:\footnote{Instead of a quasi-Poisson error model for DLNM, we fit a Gaussian error model as a reasonable approximation.}
\begin{equation}
    \hat{\bm{e}}^{(0)}_x = \mathcal{S}_x^{(0)}\left(U_{\tau_t}, \dots, U_{\tau_t-L}, \text{HWD}_t, \text{CWD}_t\right), \ \text{for} \,\,x = 1, \dots, N.
\end{equation}
After that, fitted DLNM components are removed from the log mortality matrix:
\begin{equation}
    \log \hat{\bm{m}}^{(1)} = \log \hat{\bm{m}}^{(0)} - \hat{\bm{e}}^{(0)},
\end{equation}
where $\log \hat{\bm{m}}^{(1)}$ is the new log mortality matrix for next recursion. We then fit the Lee--Carter model on $\log \hat{\bm{m}}^{(1)}$ and obtain the partial residuals $\hat{\bm{e}}^{(1)} = [\bm{e}^{(1)}_1, \dots, \bm{e}^{(1)}_N ]$. The partial residuals $\hat{\bm{e}}^{(1)}$ are fitted by the DLNM separately by each age group $x$. By repeating these steps for $\log \hat{\bm{m}}^{(j)}$ and $\hat{\bm{e}}^{(j)}$ with $j \geq 1$, we obtain the parameter set $\bm{\Theta}_{\text{LC}}$ in the $j^{th}$ recursion $\bm{\Theta}_{\text{LC}}^{(j)}$, and the $(j+1)^{th}$ log mortality matrix:
\begin{equation}
    \log \hat{\bm{m}}^{(j+1)} = \log \hat{\bm{m}}^{(j)} - \hat{\bm{e}}^{(j)}.
\end{equation}
The algorithm stops either when the following converging condition is satisfied:
\begin{equation}
\sup_{\bm\theta\in\bm{\Theta}_{\text{LC}}}\Vert\bm\theta^{(j)} - \bm\theta^{(j-1)}\Vert_{\infty} < \delta,
\end{equation}
or when the maximum number of iterations $J$ is reached. The algorithm is detailed in Algorithm \ref{bdlnmlc}.

\begin{algorithm}[H]
\label{bdlnmlc}
\caption{Backfitting DLNM--LC}
\setstretch{1.2}
\KwInput{An $N \times T$ (age $\times$ time) mortality matrix $\bm{m}$}
\KwOutput{$\log (\hat{\bm{m}}^{(r)})$ for $t = 1,2, \dots, T$, $\hat{\bm{\kappa}}^{(r)}$, $\hat{\bm{a}}^{(r)}$, and $\hat{\bm{b}}^{(r)}$, $ r \in [1, J]$.} 
\KwProcedure{DLNM--LC($\bm{m}$)}
$\hat{\bm{m}}^{(0)} \leftarrow \bm{m}$

$j \leftarrow 0$  

\While{$j \leq J$}{
$\bm{\hat{a}}^{(j)}, \hat{\bm{b}}^{(j)}, \hat{\bm{\kappa}}^{(j)} \leftarrow LC(\hat{\bm{m}}^{(j)})$

${\bm{e}}^{(j)} \leftarrow \log \hat{\bm{m}}^{(j)} - \hat{\bm{a}}^{(j)}$

Decompose ${\bm{e}}^{(j)} = \left[{\bm{e}}_1^{(j)}, \dots, {\bm{e}}_N^{(j)}\right]$ by age group $x$

Fit $\hat{\bm{e}}_x^{(j)} = \mathcal{S}_x^{(j)}\left(U_{\tau_t}, \dots, U_{\tau_t-L}, \text{HWD}_t, \text{CWD}_t\right), \ \forall x$

$\log{\hat{\bm{m}}}^{(j+1)} \leftarrow\log {\hat{\bm{m}}^{(j)}} - \hat{\bm{e}}^{(j)}$

\If{$j = J$ \textnormal{\textbf{or}}
    \textnormal{(}$j > 0$ \textnormal{\textbf{and}}
     $\sup_{\theta\in\bm{\Theta}_{\mathrm{LC}}}\Vert\theta^{(j)} - \theta^{(j-1)}\Vert_\infty < \delta$\textnormal{)}}{
  $r \leftarrow j$
  
  \textbf{break}
}

\Else{$j \leftarrow j+1$}
}
\end{algorithm}
\vspace{0.5em}
In this study, we set the maximum number of iterations $J = 20$. The algorithm stops if the change in each parameter is less than $\delta = 10^{-2}$. The fitted mortality rates under log scale are obtained as follows:
\begin{equation}
\log\hat{m}(x,t) = \hat{a}^{(r)}(x) + \hat{b}^{(r)}(x)\hat{\kappa}^{(r)}(t) + \sum_{j=0}^{r-1} \mathcal{S}_x^{(j)}\left(U_{\tau_t}, \dots, U_{\tau_t-L}, \text{HWD}_t, \text{CWD}_t\right),
\end{equation}
where the Lee--Carter component is obtained when the algorithm is converged in the $r^{th}$ recursion. 
The climate-driven mortality component is estimated by the sum of all DLNM components across $(r-1)$ iterations. 
The sum over DLNM-components appears, because expanding the steps $\log{\hat{\bm{m}}}^{(j+1)} \leftarrow\log {\hat{\bm{m}}^{(j)}} - \hat{\bm{e}}^{(j)}$ for $j = 0, \dots, r - 1$ gives 
   $\log{\bm{m}} = \log{\hat{\bm{m}}^{(0)}} = \log{\hat{\bm{m}}}^{(r)} + \sum_{j = 0}^{r - 1} {\hat{\bm{e}}}^{(j)},$
where $\hat{\bm{e}}^{(j)}$ is the $j$-th DLNM component and $\log{\hat{\bm{m}}}^{(r)}$ the LC model in the $r$-th iteration.

However, this climate-driven mortality component can be equivalently expressed as a single DLNM. 
More specifically, the sum of the coefficients obtained from the multiple DLNM iterations are equal to the coefficients of this single DLNM. The detailed explanation is provided in online supplementary material Section B.

Once the DLNM--LC model is fitted, we apply it to conduct mortality forecasting.
In this study, we consider both point forecast of mortality rates with observed UTCI, as well as scenario-based forecast of mortality rates with UTCI generated from climate models. 
First, we apply a time series approach to forecast the stochastic mortality component. The optimal ARIMA model, with potential weekly seasonal lags of 52, is selected and estimated using an R package \texttt{forecast} \citep{hyndman2020forecast}.\footnote{Details of the selected optimal time series models are presented in Tables D.1 and D.2 of the online supplementary material Section D, for both DLNM--LC and DLNM--LL models.}
Second, we obtain the point forecast of DLNM component via fitted parameters $\hat{\bm{\zeta}}^{(j)}$ with $j = 0, \dots, r-1$ in summed DLNMs. For interval forecast of the DLNM component, we incorporate parameter uncertainty using the bootstrapping method described in Section \ref{sec:2.1}. 

\subsection{Multi-population setting: the DLNM--LL model}
\label{sec:Multi-population setting: DLNM--LL model}
Following the popularity of the Lee--Carter model, \cite{li2005coherent} proposed a multi-population extension, commonly referred to as the Li--Lee model. 
The model assumes that similar regions share a common long-term mortality improvement trend, supplemented by region-specific components.
The model is formulated as follows:

\begin{equation}
\log m(x,t,i) = A(x,i) + B(x)K(t) + b(x, i)\kappa(t, i) + \epsilon(x,t,i),
\end{equation}
for $x = 1, \dots, N$, $t=1, \dots, T$, and $i = 1, \dots, n$, where $A(x,i)$ represents the age-specific mean of mortality rate over time for region $i$, $K(t)$ represents the common time trend in mortality across all regions, and $B(x)$ is the age-specific loading associated with $K(t)$, $\kappa(t, i)$ represents the region-specific mortality trend for region $i$ in addition to the common trend, and $b(x,i)$ is the corresponding age-specific loading. $\epsilon(x,t,i)$ is the error term of the model. Similar to the Lee--Carter model, to solve the model identification issue, the following constraints are imposed:
\begin{equation}
\sum_{x=1}^N B^2(x) = 1  \ \text{and} \ \sum_{t=1}^T K(t) = 0,
\end{equation}
\begin{equation}
\sum_{x=1}^N b^2(x, i) = 1  \ \text{and} \ \sum_{t=1}^T \kappa(t, i) = 0, \ \text{for} \ i=1,\dots, n.
\end{equation}

To estimate the Li--Lee model, we follow the product-ratio functional method proposed in Hyndman--Booth--Yasmeen (HBY) model \citep{hyndman2013coherent}. The HBY model can be considered a generalization of the Li–Lee model, allowing for additional bilinear age–time interaction terms to be included.
The HBY model decomposes mortality rates into a product term and a ratio term, applying SVD to retain up to six principal components. If we choose to retain only the first principal component for both the product term and the ratio term, the HBY model simplifies to the Li--Lee model.

The detailed estimation procedure of the Li--Lee component is outlined as follows. 
First, let $\bm{M}$ denote an $N \times T\times n$ (age × time × region index) mortality tensor with entry $m(x,t,i)$. 
We then decompose the mortality rate as $m(x,t,i) = p(x,t)r(x,t,i)$, where $p(x,t)$ represents the product term, calculated as the geometric mean of mortality rate across each region, and $r(x,t,i)$ represents the ratio term of region $i$'s mortality rate to $p(x,t)$.
We fit separate Lee--Carter models to $p(x,t)$ and $r(x,t,i)$, yielding common factors $\hat{\bm{A}}_p$, $\hat{\bm{B}}$, $\hat{\bm{K}}$ and region-specific factors $\hat{\bm{a}}$, $\hat{\bm{b}}$, $\hat{\bm{\kappa}}$, respectively. 
We then compute the age-specific mean as $\hat{\bm{A}} = \hat{\bm{A}}_p + \hat{\bm{a}}$.
This is summarized in online supplementary material Section A. 

We define the estimated parameter set of the Li–Lee model as follows: $\bm{\Theta}_{\text{LL}}^{(j)} := \{\hat{\bm{A}}, \hat{\bm{B}}, \hat{\bm{K}}, \hat{\bm{b}}, \hat{\bm{\kappa}}\}$.
We integrate the DLNM component into the Li--Lee model to incorporate climate impacts. The proposed DLNM--LL model is specified as follows:
\begin{align}
\label{dlnmll_expression}
\log m(x,t,i) &= \underbrace{A(x,i) + B(x) K(t) + b(x,i) \kappa(t,i)}_{\text{Li--Lee component}} \nonumber \\
&+ \underbrace{\mathcal{S}_{x,i}\left(U_{\tau_{t,i}}, \dots, U_{\tau_{t-L,i}}, \text{HWD}_{t,i}, \text{CWD}_{t,i}\right)}_{\text{DLNM component}} + \epsilon(x,t,i),
\end{align}
where $A(x,i)$, $B(x)$, $K(t)$, $b(x,i)$ and $\kappa(t,i)$ represent the Li--Lee model parameters. The term $\mathcal{S}_{x,i}\left(U_{\tau_t}, \dots, U_{\tau_t-L}, \text{HWD}_t, \text{CWD}_t\right)$ characterizes the non-linear relationship between UTCI and mortality rates for age group $x$ and region $i$.
$\epsilon(x,t,i)$ is the error term of the model. The Li--Lee component captures the inherent mortality trends, including the common trend and region-specific trend, while the integrated DLNM component isolates and quantifies the climate-driven mortality risks for each region and age group.

To estimate the DLNM--LL model, again we adopt the backfitting approach and jointly estimate the two model components. We denote the original mortality tensor $\log \bm{M} = \log \hat{\bm{M}}^{(0)}$, where 

\begin{equation}
    \log {\bm{M}} = \left[\log {\bm{m}}^{(0)}(x,t,1), \dots, \log {\bm{m}}^{(0)}(x,t,n) \right].
\end{equation}
The estimation procedure is similar to the estimation of DLNM--LC model. For the stochastic mortality component, we fit the Li--Lee model on the mortality tensor. For the DLNM components, we fit a DLNM on each age group in each region. The estimation details are described in online supplementary material A, and the algorithm of DLNM-LL model is summarized in Algorithm \ref{bdlnmll}.

\vspace{0.1in}
\begin{algorithm}[H]
\label{bdlnmll}
\caption{Backfitting DLNM--LL}
\setstretch{1.2}
\KwInput{An $N \times T \times n$ mortality tensor $\bm{M} = [\bm{m}(x,t,1), \dots, \bm{m}(x,t,n)]$}
\KwOutput{$\hat{\bm{A}}^{(r)}$, $\hat{\bm{B}}^{(r)}$, $\hat{\bm{K}}^{(r)}$, $\hat{\bm{b}}^{(r)}$, $\hat{\bm{\kappa}}^{(r)}$, and $\hat{\bm{M}}^{(r)}$, $r \in [1,J]$}
\KwProcedure{DLNM--LL($\bm{M}$)}

$\hat{\bm{M}}^{(0)} \leftarrow \bm{M}$

$j \leftarrow 0$

\While{$j \leq J$}{

$\hat{\bm{A}}^{(j)}, \hat{\bm{B}}^{(j)}, \hat{\bm{K}}^{(j)}, \hat{\bm{b}}^{(j)}, \hat{\bm{\kappa}}^{(j)} \leftarrow LL(\hat{\bm{M}}^{(j)})$

$\bm{e}^{(j)}_i \leftarrow \log\hat{\bm{m}}^{(j)}(x,t,i) - \hat{\bm{A}}^{(j)}(x,i), \ \forall i$

Decompose $\hat{\bm{e}}_i^{(j)} = \left[\hat{\bm{e}}_{1,i}^{(j)}, \dots, \hat{\bm{e}}_{N,i}^{(j)}\right]$ by age group $x$

Fit $\hat{\bm{e}}_{x,i}^{(j)} = \mathcal{S}_{x,i}^{(j)}\left(U_{\tau_{t,i}}, \dots, U_{\tau_{t,i}-L}, \text{HWD}_{t,i}, \text{CWD}_{t,i}\right), \ \forall x,i$

$\hat{\bm{e}}^{(j)}  \leftarrow \left[\hat{\bm{e}}^{(j)}_{1,i}, \dots, \hat{\bm{e}}^{(j)}_{N,i} \right]$

$\log(\hat{\bm{M}}^{(j+1)}) \leftarrow \log(\hat{\bm{M}}^{(j)}) - \hat{\bm{e}}^{(j)}$

\If{$j = J$ \textnormal{\textbf{or}}
    \textnormal{(}$j > 0$ \textnormal{\textbf{and}}
     $\sup_{\bm\theta\in\bm{\Theta}_{\text{LL}}}\Vert\bm\theta^{(j)} - \bm\theta^{(j-1)}\Vert_\infty < \delta$\textnormal{)}}{
  $r \leftarrow j$
  
  \textbf{break}
}

\Else{$j \leftarrow j+1$}
}
\end{algorithm}
\vspace{0.1in}

The maximum number of iterations $J$ is set again at 20.
The algorithm stops if the change in each parameter is less than $\delta = 10^{-2}$.
The fitted log mortality rates are obtained as follows:
\begin{align}
\label{dlnmll_fitted}
\log\hat{m}(x,t,i) &=\hat{A}^{(r)}(x, i) + \hat{B}^{(r)}(x)\hat{K}^{(r)}(t) + \hat{b}^{(r)}(x,i)\hat{\kappa}^{(r)}(t,i) \nonumber \\
&+ \sum_{j=0}^{r-1} \mathcal{S}_{x,i}^{(j)}\left(U_{\tau_{t,i}}, \dots, U_{\tau_{t,i}-L}, \text{HWD}_{t,i}, \text{CWD}_{t,i}\right).
\end{align}
For the Li--Lee component, we fit separate time series models to $K(t)$ and $\kappa(t,i)$. As demonstrated in online supplementary material Section B, the summed DLNM components can be represented by a single DLNM.
We generate mortality forecasts using the DLNM--LL model under the same setting as in Section \ref{sec:Single-population setting: DLNM--LC model}.

\subsection{Quantification of climate-driven mortality risk}
Under the proposed modeling framework in Sections \ref{sec:Single-population setting: DLNM--LC model} and \ref{sec:Multi-population setting: DLNM--LL model}, mortality time series can be decomposed into a stochastic component and a climate-driven component. In this section, we introduce and define a climate loading measure to quantify climate-driven mortality risk relative to total mortality.

First, we denote $\tilde{m}$ as the estimated stochastic mortality component, which is calculated as: 
\begin{equation}
\log \tilde{m}(x,t,i) =\log \hat{m}(x,t,i)-\sum_{j=0}^{r-1} \mathcal{S}_{x,i}^{(j)}\left(U_{\tau_{t,i}}, \dots, U_{\tau_{t,i}-L}, \text{HWD}_{t,i}, \text{CWD}_{t,i}\right).
\end{equation}

The climate loading ${\theta}$ is then defined as:

\begin{equation}
{\theta}(x,t,i) = 1-\frac{\tilde{m}(x,t,i)}{\hat{m}(x,t,i)}=\frac{\hat{m}(x,t,i) - \tilde{m}(x,t,i)}{\hat{m}(x,t,i)}.
\label{theta}
\end{equation}

${\theta}$ can be interpreted as the proportion of climate-driven mortality relative to the total mortality. Alternatively, $\frac{1}{1-\theta}$ can be considered as a climate-related risk adjustment factor applied to the stochastic mortality component $\tilde{m}$.
It should be noted that ${\theta}$ can take both positive and negative values. Under favorable weather conditions, we experience a reduction in mortality relative to the stochastic mortality component. 
Therefore, we should expect a seasonal pattern in climate loading $\theta$, \textit{i.e.} high in winter and low in summer. The absolute value of $\theta$ reflects the magnitude of climate-driven mortality relative to the total mortality. 

\section{Data}
\label{sec:Data}
\subsection{Mortality data}
We collect weekly death count data and annual population data from the Eurostat database, over the investigation period 2015--2019, for four age groups 20--64, 65--74, 75--84 and 85+, and in three different NUTS\footnote{NUTS stands for the Nomenclature of Territorial Units for Statistics.} regions: Attica (EL30, referred to as Athens\footnote{As the population of Attica is largely concentrated in Athens, we referred to the region simply as Athens.}), \'Area Metropolitana de Lisboa (PT170, referred to as Lisbon), and Roma (ITI43, refer to as Rome).\footnote{Data source: \url{https://ec.europa.eu/eurostat/databrowser/view/demo_r_mweek3/default/table?lang=en}. Note that weekly death data across different age groups at the NUTS 3 level are only available from 2015 onwards.} We choose these three regions due to their similar demographic characteristics and Mediterranean climate conditions.

\begin{figure}[H]
    \centering
\includegraphics[width=1\linewidth]{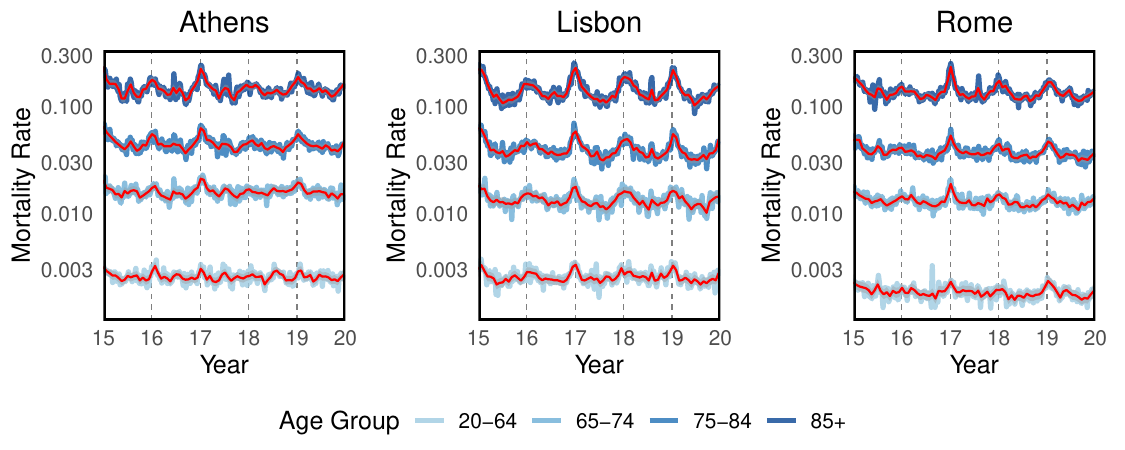}
\vspace{-0.2in}
    \caption{Historical weekly mortality rates (2015--2019).}
    \label{fig:hist_mort}
\end{figure}

Let $D(x,t,i)$ be the death count for age $x$, region $i$ at time $t$, and $P(x,t,i)$ for the corresponding population size. 
We assume that the population size is constant throughout the year.  
We also assume that there are 52 weeks in each year, so the weekly risk exposure can be approximated as $E(x,t,i) = P(x,t,i)/52$, the weekly mortality rate is thus calculated as:
\begin{equation}
m(x,t,i) = \frac{D(x,t,i)}{E(x,t,i)} :=  \frac{D(x,t,i)}{P(x,t,i)/52}.
\end{equation}
Figure \ref{fig:hist_mort} visualizes weekly mortality data from 2015 to 2019 for the three regions considered in this study. Overall, the regions exhibit consistent trends in mortality rates across age groups: older age groups consistently show higher mortality rates and stronger seasonal fluctuations compared to younger age groups. Interestingly, in addition to the well-known cyclical pattern (higher mortality in winter and lower mortality in summer), we detected a W-shaped seasonal pattern in certain years. 
This W-shaped pattern consists of two distinct peaks: the first and larger peak occurs during winter, driven by the usual increase in mortality, while the second and smaller peak arises in summer, caused by an unexpected rise in mortality due to extreme heat waves. This W-shaped pattern is particularly apparent in the age groups 75--84 and 85+, and can be observed across all three regions. Due to the strong seasonality in the weekly mortality data and the relatively short observation period, it is difficult to detect any mortality trend through simple visual examination.

\subsection{UTCI data}
The UTCI is defined as the air temperature $(^\circ\text{C})$ under standard reference conditions that would elicit the same physiological response\footnote{Examples of physiological responses include sweat production, shivering, skin wetness, skin blood flow, and mean skin and face temperatures.} as the actual environmental conditions, which depend on factors such as wind speed, humidity, and radiation \citep{blazejczyk2013introduction}. It reflects perceived temperature rather than the actual air temperature, accounting for thermal stress on the human body \citep{fiala2012utci}.\footnote{We note that previous research, such as \cite{guo2024regional} and \cite{lo2023optimal}, has found dry-bulb temperature to be one of the best predictors of mortality. In this paper, UTCI is used as an illustrative metric to demonstrate the proposed methods, and there is no evidence that UTCI predicts mortality better than dry-bulb temperature. We leave the investigation of alternative temperature variables to future research.} The UTCI, denoted by $U$,  is calculated as: 
\begin{equation}
\label{utci_eq}
U = T_a + f(T_a, \text{RH}, \text{WS}, \text{MRT}),
\end{equation}
where $T_a$ is 2m air temperature, RH is near-surface relative humidity, WS is 10m wind speed, MRT is mean radiant temperature, and $f(.)$ is a $6^{th}$ order polynomial approximation function.
In this study, hourly UTCI data spanning from 2015 to 2019, with a grid resolution of $0.1^{\circ} \times 0.1^{\circ}$ based on ERA-5 Reanalysis, are collected from the Climate Data Store\footnote{Data source: \url{https://cds.climate.copernicus.eu/datasets/reanalysis-era5-single-levels?tab=download}.}. Nearest grid areas are selected for each region's center to acquire hourly UTCI data. These hourly data are then used to produce daily mean, minimum, and maximum UTCI. 

\begin{figure}[H]
\centering
\includegraphics[width=0.8\linewidth]{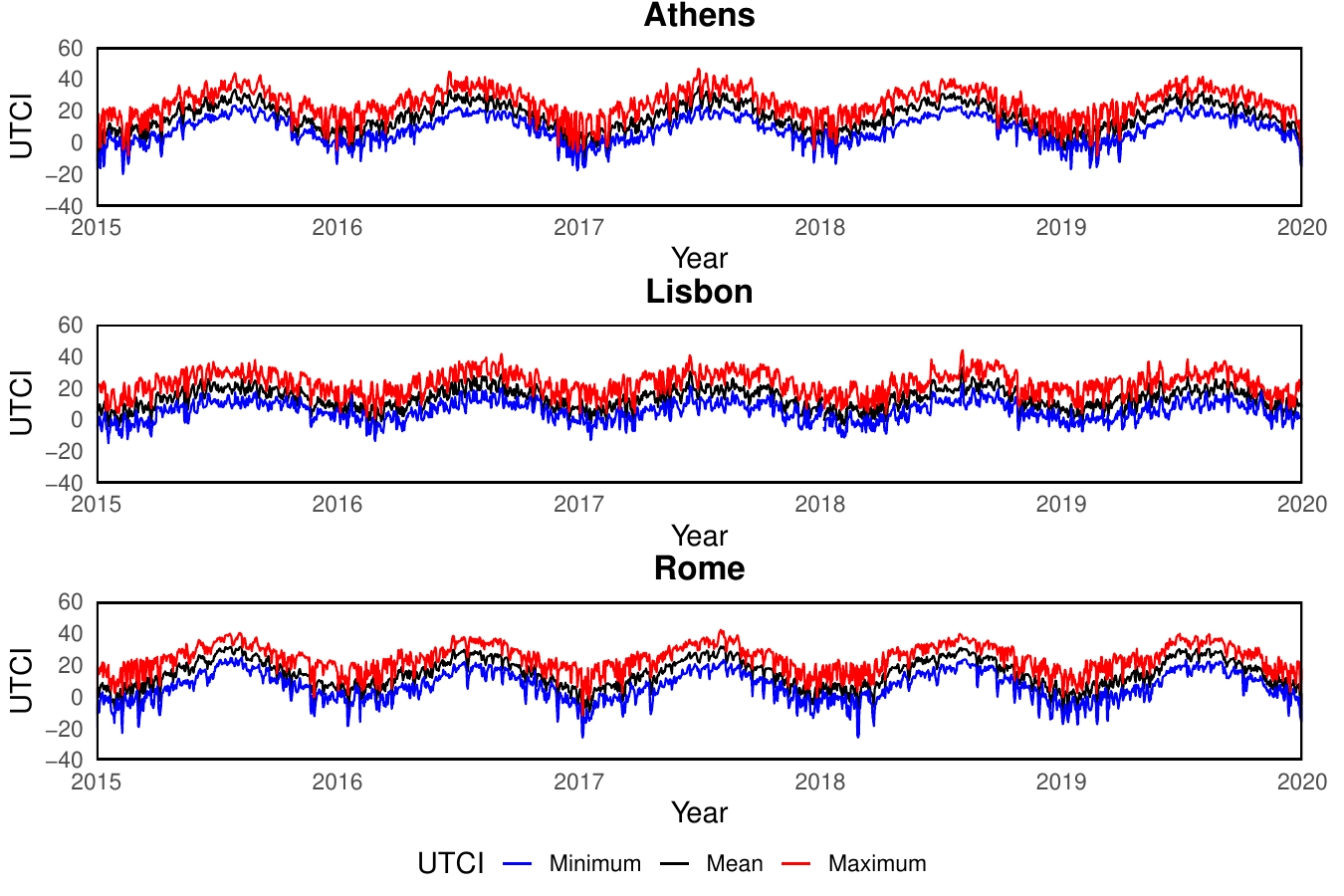}
\vspace{-0.1in}
\caption{Historical daily UTCI (2015--2019).}
\label{utci}
\end{figure}

In Figure \ref{utci}, we plot the daily UTCI during 2015--2019 for the Athens, Lisbon, and Rome. We can see that the three regions have similar UTCI levels and seasonal patterns. The UTCI in Rome shows slightly stronger level of cold stress in winter compared to the other two regions.
Due to the nature of the Mediterranean climate, cold stress in the three regions is generally minimal, with cold waves being rare events during the 2015--2019 period. 
On the other hand, heat stress remains a major threat to population health in Southern Europe. For example, during the ``Lucifer'' heatwave in the summer of 2017, record-breaking temperatures were observed, with locations including Rome reaching over 48$^\circ\text{C}$. As a result, hospitals across the affected regions experienced a 15\% increase in admissions due to heat-related illnesses \citep{kew2019lucifer}. From Figure \ref{utci}, we observe noticeable spikes in the maximum UTCI during the summer seasons of 2017--2018.

\subsection{Heatwave and coldwave variables}
\label{sec:Heat wave and cold wave variable}
In addition to daily mean UTCI, we also want to incorporate information of maximum and minimum temperatures into the model. In doing so, we include heatwave and coldwave variables to capture the short-term mortality increase resulting from prolonged periods of extreme temperatures. We therefore introduce the number of heat wave days ($\text{HWD}_t$) and cold wave days ($\text{CWD}_t$) in week $t$ as part of the DLNM component.\footnote{We also consider alternative definitions of HWD and CWD variables using a percentile-based approach  \cite[see e.g.][]{guo2017heat}. The main conclusions of the study remain unchanged. These additional results are available in online supplementary material Section E.} 
These variables account for the non-additive effects of consecutive extreme heat and cold days. 
Moreover, they are calculated using daily maximum and minimum UTCI data, respectively, fully utilizing the UTCI data to capture extreme climate conditions. 
UTCI values can be classified into ten heat and cold stress levels \citep[][]{di2020thermal}. Under this classification, 32$^\circ\text{C}$ is the threshold for strong heat stress, and $-13^\circ\text{C}$ is the threshold for strong cold stress.
In this way, we identify a heatwave or coldwave day if the daily maximum UTCI is above 32, or daily minimum UTCI is below $-13$, for three consecutive days, respectively. The variables for the number of heatwave days and coldwave days in week $t$ are then defined as:
\begin{equation}
\text{HWD}_t = \sum_{\tau = 1 + 7(t - 1)}^{7t}\mathbbm{1}\{U_{\tau}^{\max} > 32\} \times\mathbbm{1}\{U^{\max}_{\tau-1} > 32\} \times\mathbbm{1}\{U^{\max}_{\tau-2} > 32\},
\end{equation}
\vspace{-0.5em}
\begin{equation}
\text{CWD}_t = \sum_{\tau = 1 + 7(t - 1)}^{7t} \mathbbm{1}\{U_{\tau}^{\min} < -13\} \times\mathbbm{1}\{U^{\min}_{\tau-1} < -13 \} \times\mathbbm{1}\{U^{\min}_{\tau-2} < -13\},
\end{equation}
where $\mathbbm{1}(.)$ is an indicator function, $U_{\tau}^{\min}$ and $U_{\tau}^{\max}$ represent the daily minimum UTCI and the daily maximum UTCI on day $\tau$, respectively.

\section{Empirical results}
\label{sec:Empirical results}
\subsection{Stochastic mortality components}
As discussed in Section \ref{sec:Methodology}, our model separates climate-driven mortality from non-climate-driven mortality. Since the primary source of mortality seasonality is captured by the climate-driven components of the model, the stochastic mortality components in the DLNM--LC and DLNM--LL models are expected to exhibit little or no seasonality, reflecting potential non–climate-driven seasonal effects in mortality.
To validate this, we compare the time-varying factor $\kappa(t)$ obtained from the original stochastic mortality models with that of the stochastic mortality components obtained in our proposed modeling framework. 

\begin{figure}[h!]
\centering
\includegraphics[width=1\linewidth]{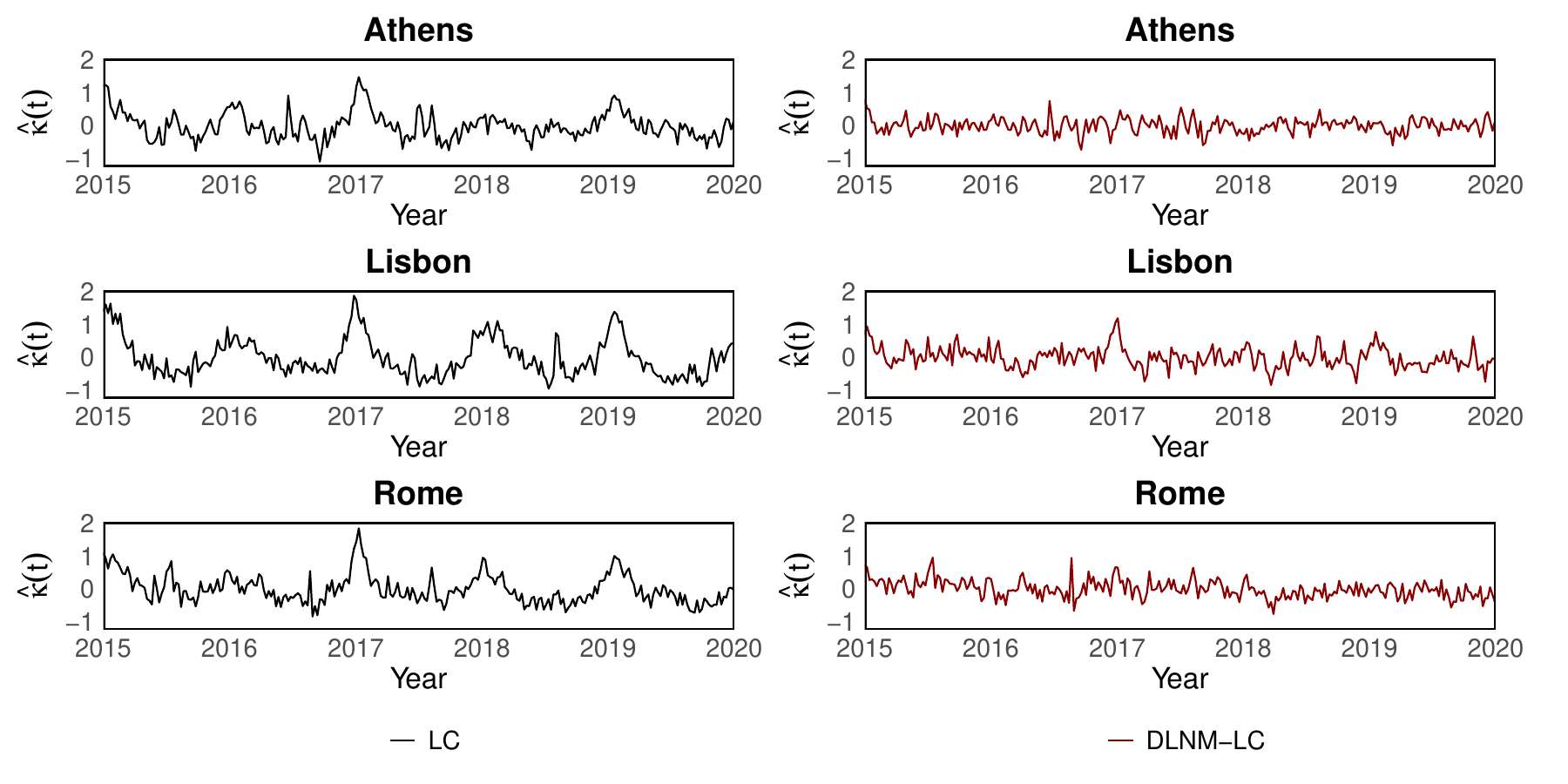}
\vspace{-0.3in}
\caption{Fitted time-varying factors in single-population models: LC (left); DLNM--LC (right).}
\label{single_k}
\end{figure}

In Figure \ref{single_k}, we plot the $\hat\kappa_t$ from the original LC on the left panel, and the $\hat\kappa_t$ from DLNM--LC on the right panel, for the three regions considered in our study. The plots on the left show very clear seasonal fluctuations, whereas the plots on the right exhibit minimal or no seasonality. 
This finding confirms that the proposed models successfully isolate climate-driven patterns in mortality from the stochastic mortality components.

\begin{figure}[h!]
\centering
\includegraphics[width=1\linewidth]{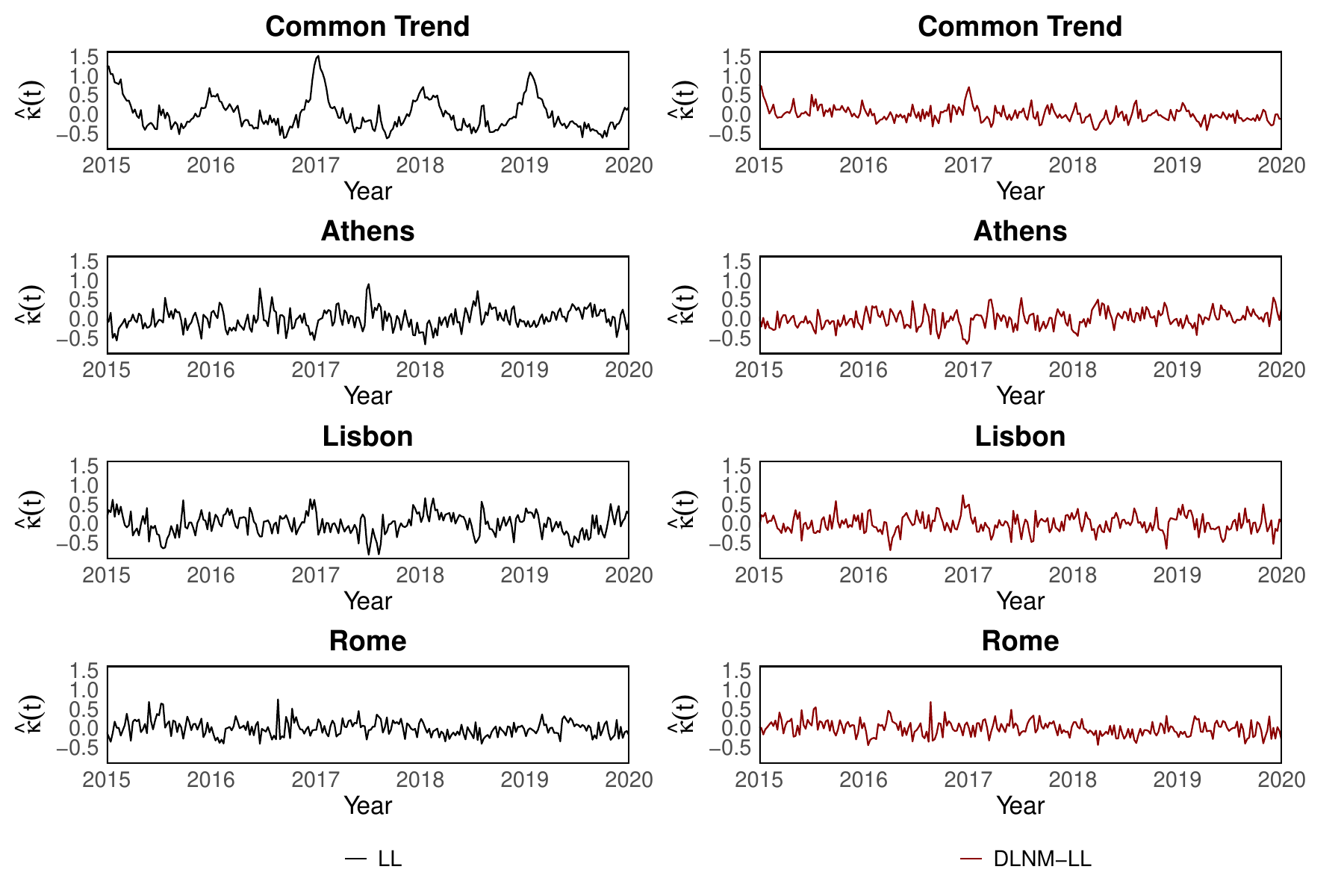}
\vspace{-0.3in}
\caption{Fitted time-varying factors in multi-population models: LL (left); DLNM--LL (right).}
\label{multi_k}
\end{figure}

Similar results can be found for the multi-population mortality models in Figure \ref{multi_k}. We plot the $\hat\kappa_t$ from the original LL on the left panel, and the $\hat\kappa_t$ from DLNM--LL on the right panel. From the LL model estimates, we can see that the common trend $K(t)$ captures most of the seasonality in the mortality data. The region-specific time trend captures any residual variations that are not explained by the common time trend, reflecting localized demographic or environmental characteristics.
The estimated common trend and region-specific trends under the DLNM-LL model exhibit significantly reduced seasonality, particularly in the common trend.
It should be noted that the seasonality of the region-specific time trend is weak in both models. Again, we conclude that the proposed models successfully isolate climate-driven seasonality from their stochastic mortality components.\footnote{We conducted formal statistical tests for potential seasonality remaining in the model  residuals and concluded that it is appropriate to assume that the residual series from the DLNM–LC and DLNM–LL models do not exhibit seasonality. These results are provided in Section D of the online supplementary material.}

\subsection{DLNM climate-driven mortality components}
\label{sec:Calibration of climate-driven mortality components via DLNM}
From Equation (\ref{dlnmll_fitted}), we identify the sum of the DLNM functions as the climate-driven component of mortality.
By the equivalence of multiple DLNMs and single DLNM under backfitting algorithm shown in online supplementary material Section B, we present the fitting results of a single DLNM for illustrative purposes. 
As the first step, we visualize the climate-driven mortality components by plotting the fitted DLNM exposure–lag–response surface, for both the DLNM--LC and DLNM--LL models.
\begin{figure}[h!]
  \centering
  \begin{subfigure}[t]{0.46\linewidth}
    \centering
    \includegraphics[width=\linewidth]{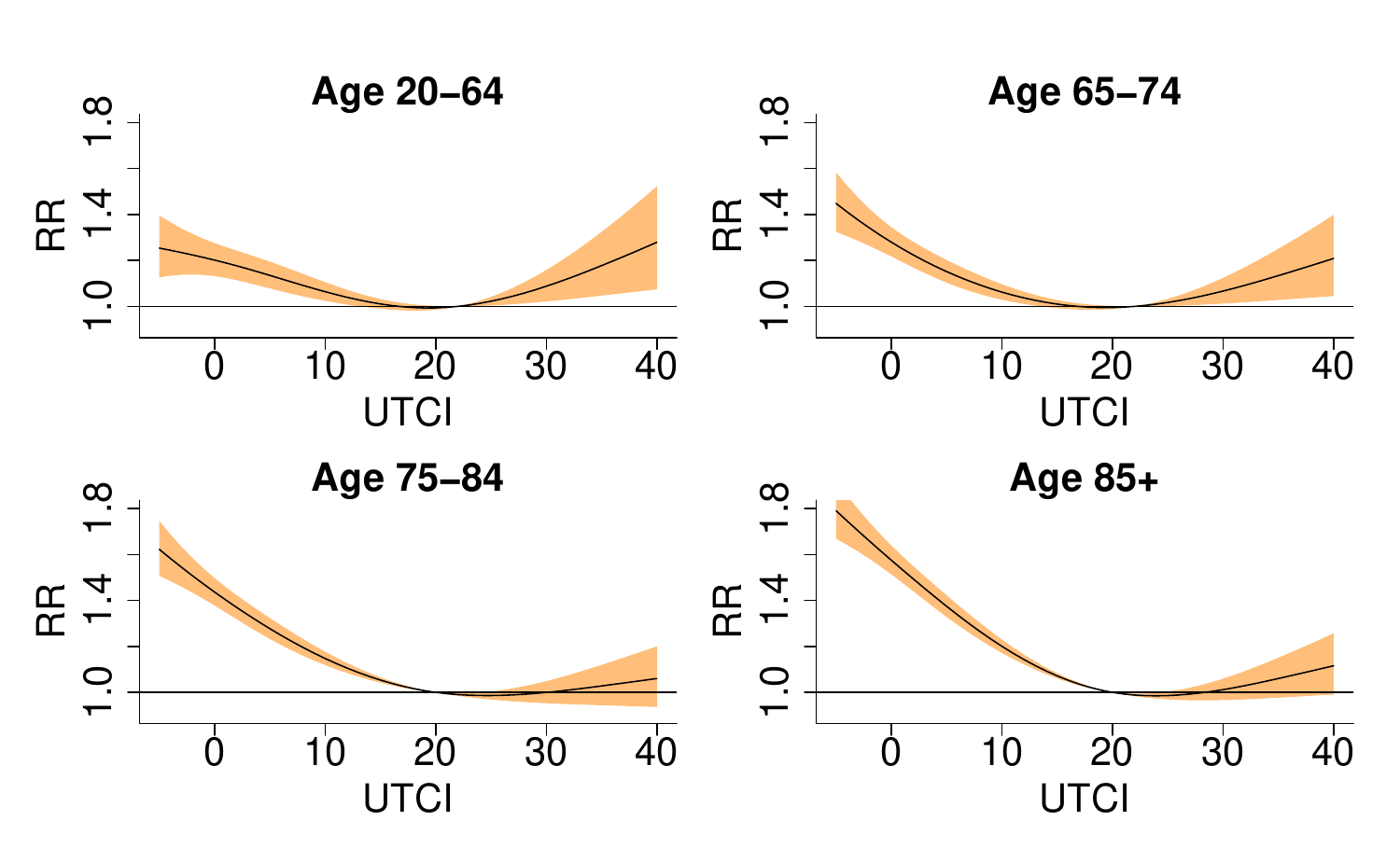}
    \subcaption{Athens, DLNM--LC model}
    \label{overall_athens_lc}
  \end{subfigure}\hfill
  \begin{subfigure}[t]{0.46\linewidth}
    \centering
    \includegraphics[width=\linewidth]{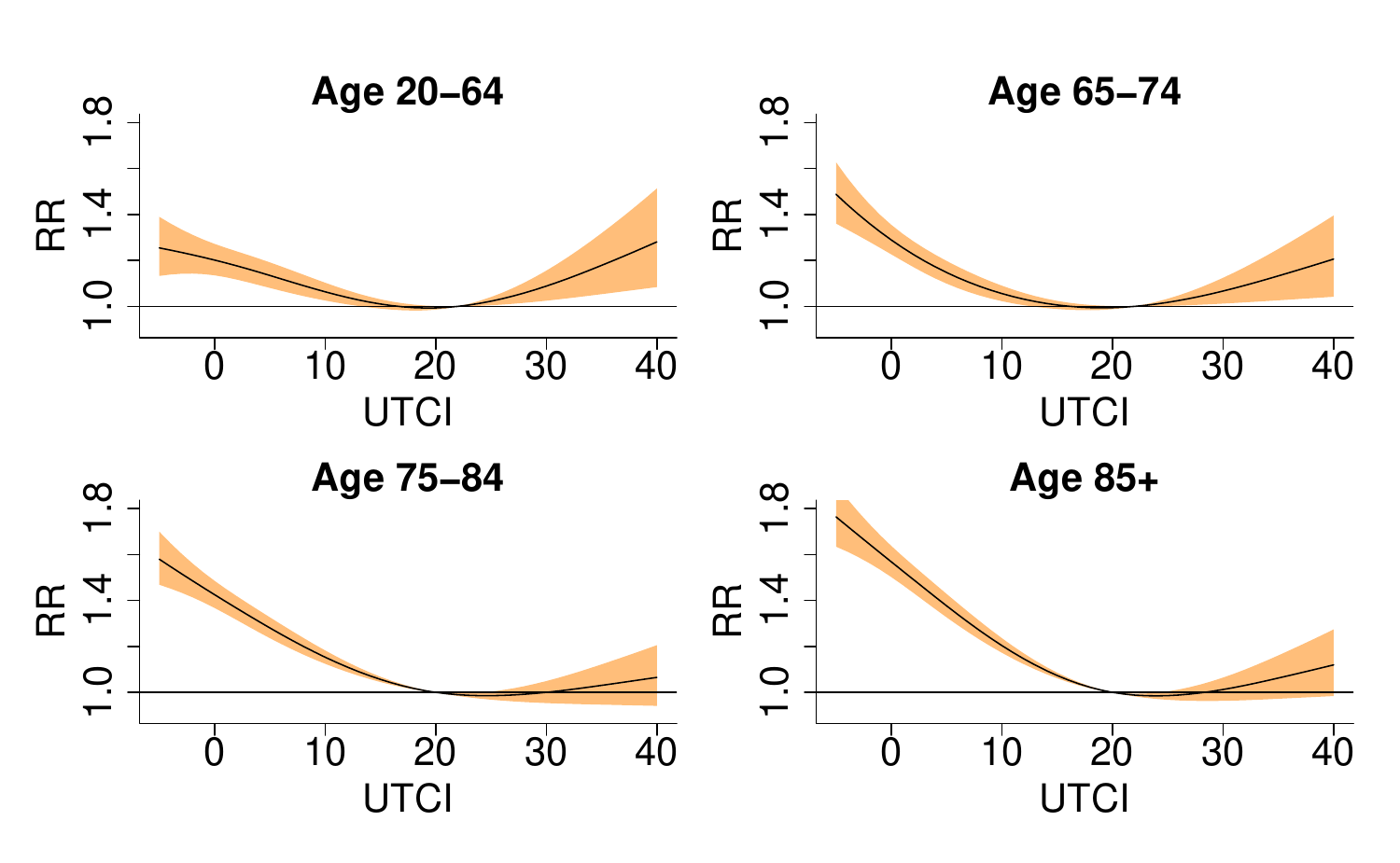}
    \subcaption{Athens, DLNM--LL model}
    \label{overall_athens_ll}
  \end{subfigure}
  \begin{subfigure}[t]{0.46\linewidth}
    \centering
    \includegraphics[width=\linewidth]{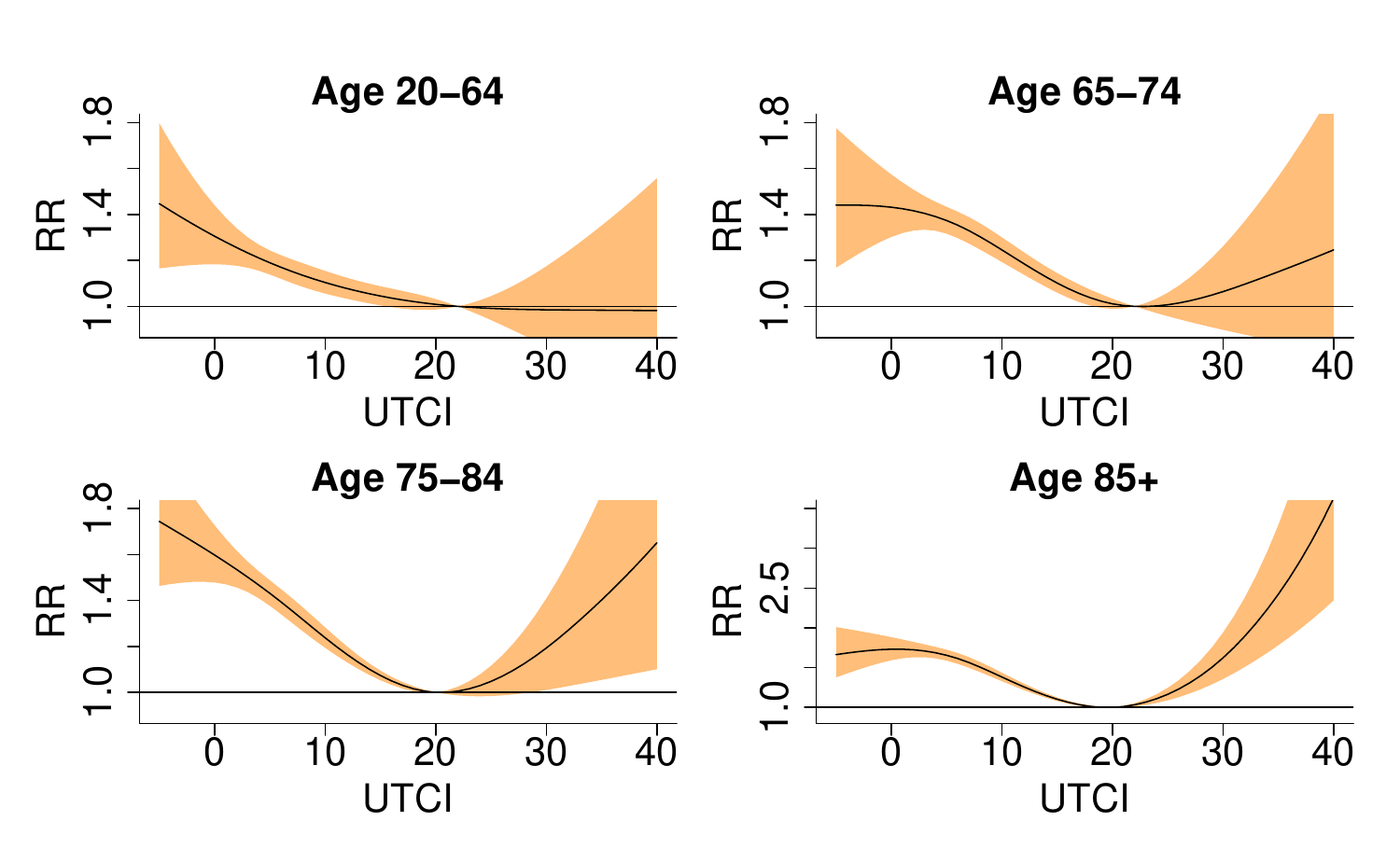}
    \subcaption{Lisbon, DLNM--LC model}
    \label{overall_lisbon_lc}
  \end{subfigure}\hfill
  \begin{subfigure}[t]{0.46\linewidth}
    \centering
    \includegraphics[width=\linewidth]{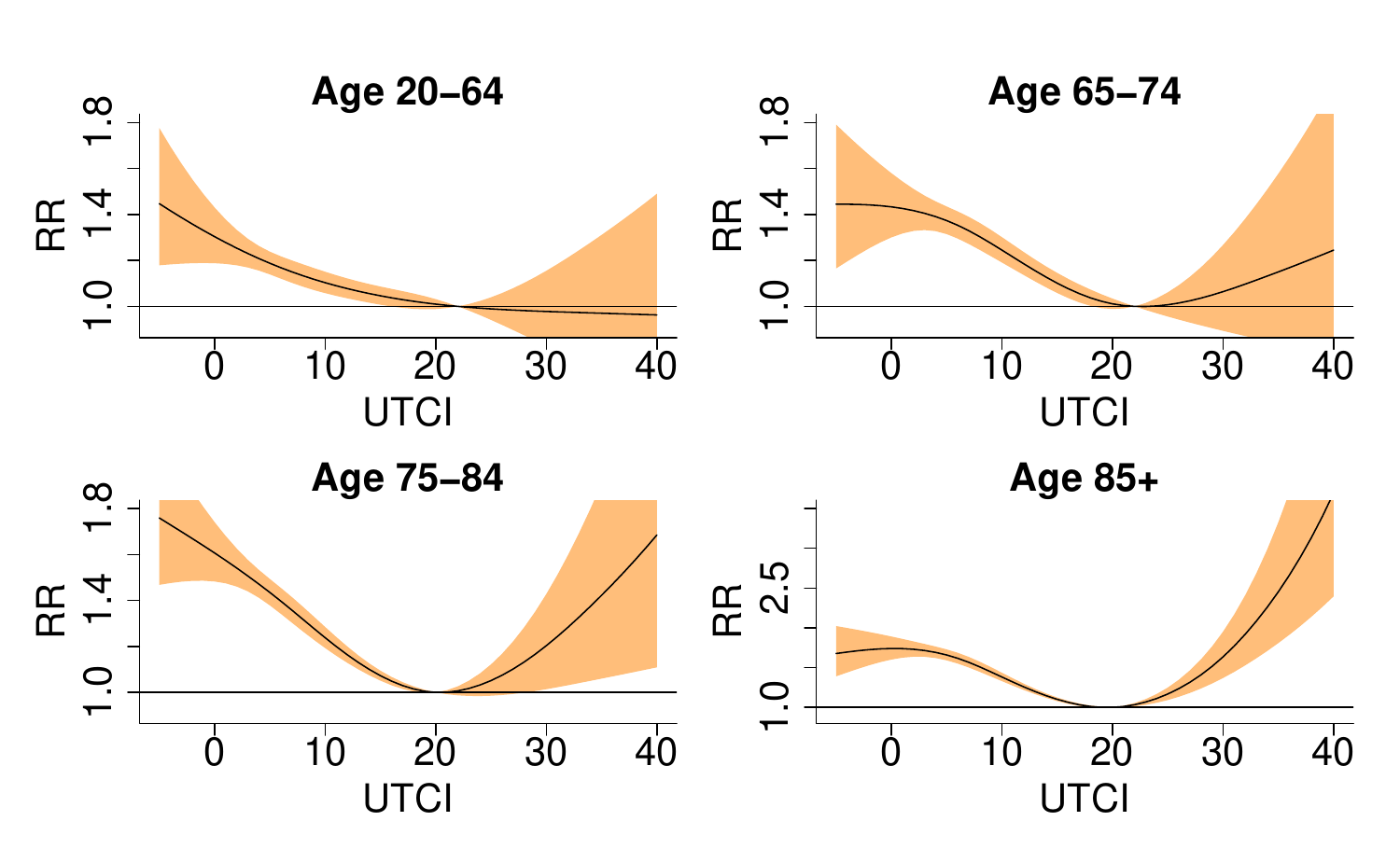}
    \subcaption{Lisbon, DLNM--LL model}
    \label{overall_lisbon_ll}
  \end{subfigure}
  \begin{subfigure}[t]{0.46\linewidth}
    \centering
    \includegraphics[width=\linewidth]{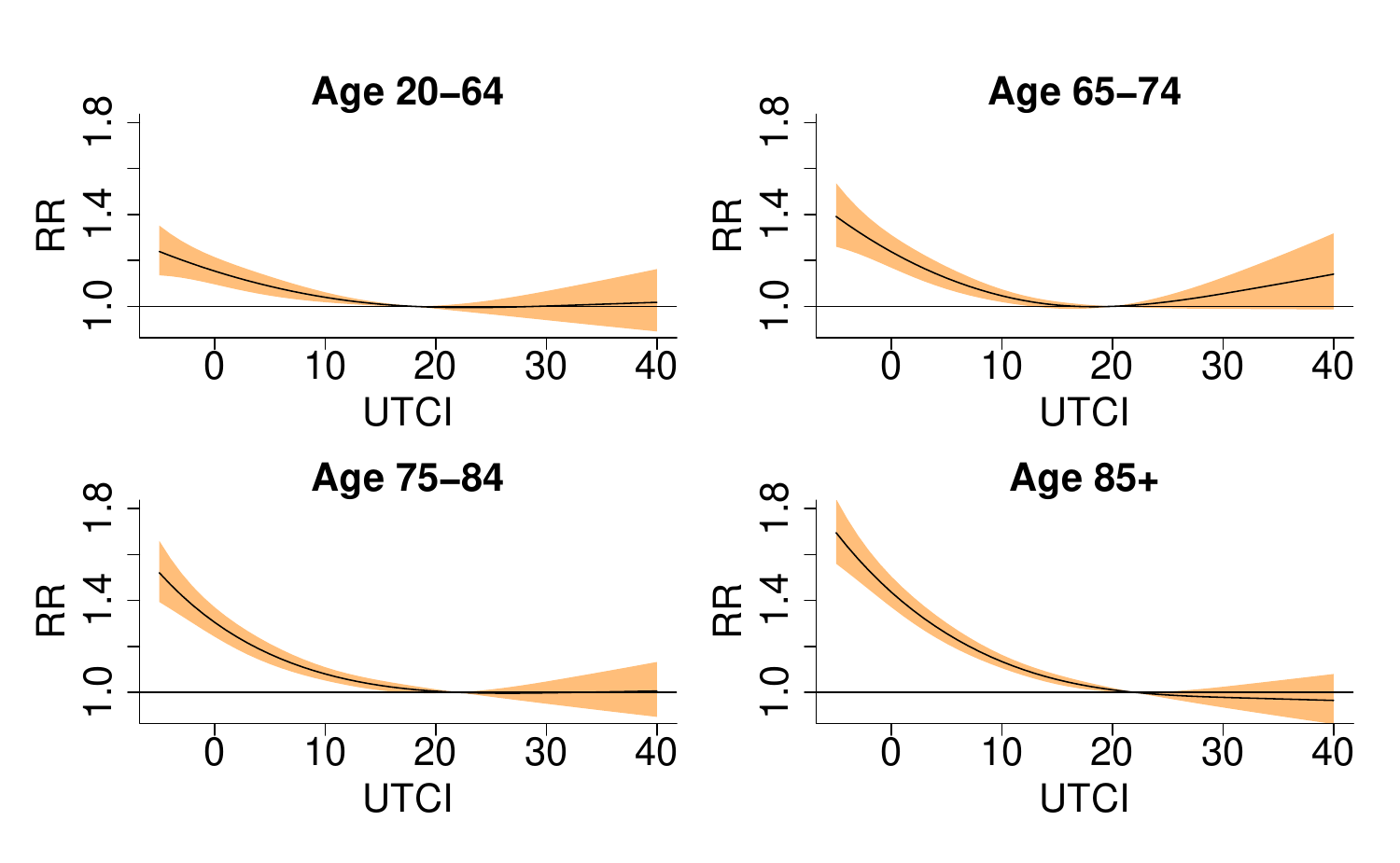}
    \subcaption{Rome, DLNM--LC model}
    \label{overall_rome_lc}
  \end{subfigure}\hfill
  \begin{subfigure}[t]{0.46\linewidth}
    \centering
    \includegraphics[width=\linewidth]{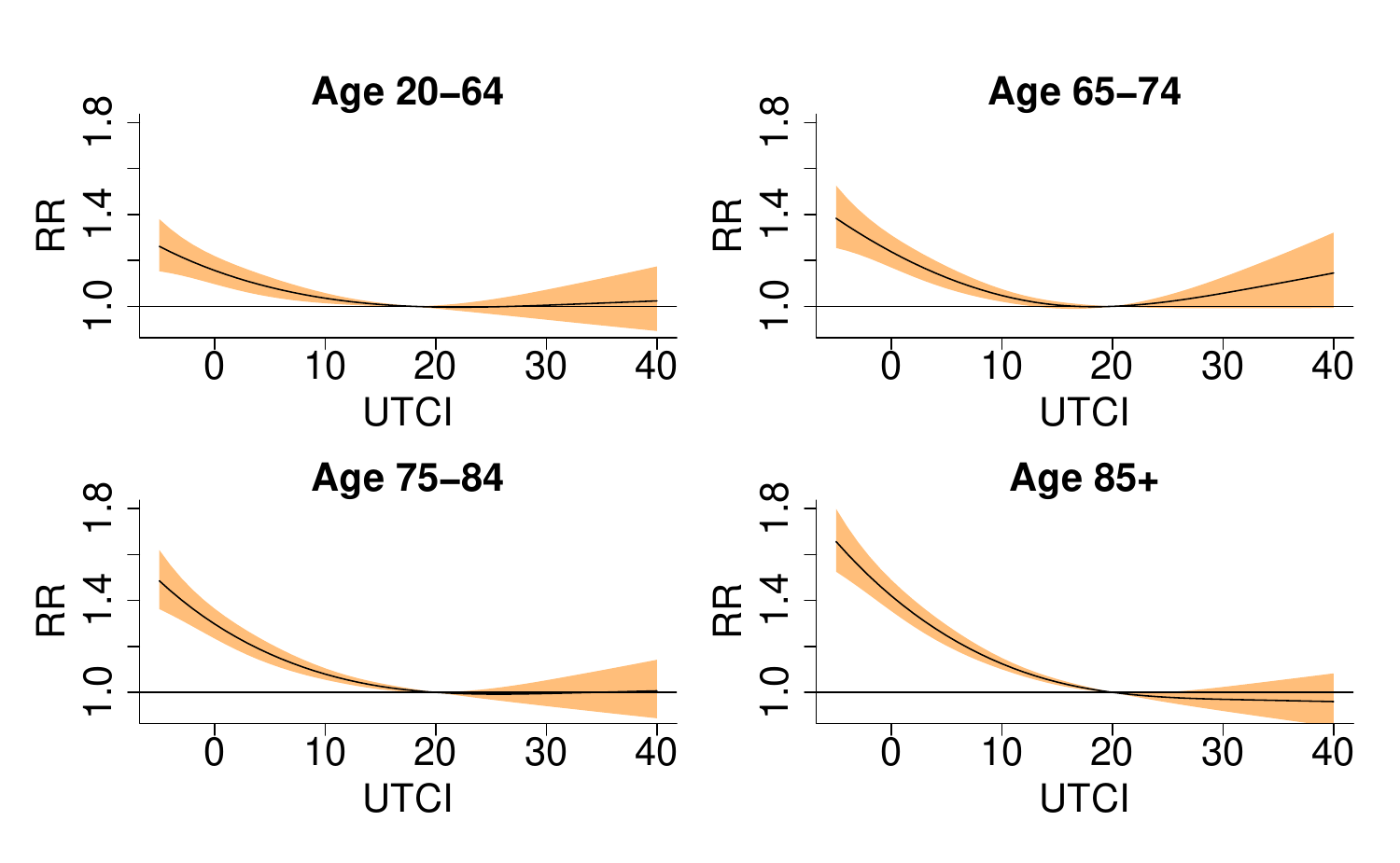}
    \subcaption{Rome, DLNM--LL model}
    \label{overall_rome_ll}
  \end{subfigure}
  \vspace{-0.05in}
  \caption{Overall cumulative effect of UTCI on climate-driven mortality components.}
  \label{fig:overall_cumulative}
\end{figure}

Figure \ref{fig:overall_cumulative} plots the overall cumulative effect of mean UTCI on climate driven mortality, and both models consistently reveal a U-shaped relationship between UTCI and the relative risk (RR) of weekly mortality. This confirms the well-established finding that both extreme cold and hot weather increase mortality risk.  
We also observe that the ``optimal'' UTCI for the minimum RR is located between approximately $18^\circ\text{C}$ and $25 ^\circ\text{C}$ in all regions. Overall, we observe that older age groups are more vulnerable to extreme cold, exhibiting higher RR at low UTCI values compared to the younger age groups. As shown in the plots, the elderly population in Lisbon is particularly sensitive to extreme heat stress across the three regions. Note that the $y$-axis scale is consistent across all panels, except for the Lisbon age group 85+, which have a higher maximum value. However, it is important to understand that the DLNM component in our models also includes the HWD and CWD variables. Therefore, to gain a complete picture of how extreme UTCI elevates mortality risk, we need to consider both the estimates from the cross-basis function and the estimated parameters for the HWD and CWD variables. This also explains why, for older age groups in Athens and Rome, the RR at high UTCI values is not as elevated as it is for younger age groups.

We plot the lagged effects of UTCI in 
Figures C.1--C.6 in online supplementary material Section C, illustrating how heat and cold exposures impact mortality over multiple lag days in the DLNM--LC and DLNM--LL models. We examine two cases where UTCI $=$ $-5$ (cold stress) and $35$ (heat stress). 
For heat exposure, the relative mortality risk typically rises within the first five days of lag, reflecting an immediate adverse effect. After around ten days of heat exposure, ``harvesting effects'' can be observed where relative risk becomes less than one. 
On the other hand, cold exposure tends to exhibit a delayed but long-lasting effect on mortality: the relative risk gradually increases over time, peaking between five and ten days of lag before subsequently declining.
These lag-response patterns are particularly pronounced in older populations.

We examine the coefficients associated with heatwave days and coldwave days in both the DLNM--LC and DLNM--LL models in Table \ref{hwd_cwd_coefficients_combined}. These results have further verified that the two models give consistent results in the DLNM components.  
We observe that across all regions and age groups, in the majority of cases, the coefficients for both $\text{HWD}_t$ and $\text{CWD}_t$ are positive and significant at the 5\% level.   
For $\text{HWD}_t$, the only exception is for age group 20--64 in Rome. For $\text{CWD}_t$, there are several cases where the coefficient is not significant at 10\%, including age groups 20--64 and 85+ in Athens, and 65--74 in Lisbon and Rome. 
Where the coefficients are significant, the results suggest that older age groups, such as those aged 75--84 and 85+, are more susceptible to mortality risks from both heat and cold extremes compared to younger age groups.
Overall, the impact of HWD is more pronounced than that of CWD. This suggests that the effect of persistent extreme heat is largely captured by the HWD variable, rather than the cross-basis function, which explains the flat portion of the U-shaped curve at high UTCI values.
We conclude that the inclusion of $\text{HWD}_t$ and $\text{CWD}_t$ complements the cross-basis matrix, ensuring that extreme risks are adequately captured by the model.\footnote{Additional analysis has been conducted on DLNM--LC and DLNM--LL models without HWD and CWD variables (see online supplementary material Section E), and we conclude that including these variables can improve overall forecasting accuracy. }

\begin{table}[H]
\centering
\renewcommand{\arraystretch}{1.2}
\resizebox{0.95\textwidth}{!}{%
\begin{tabular}{clllllll}
    \toprule
    \multirow{2}{*}{\textbf{Age group}}
      & \multirow{2}{*}{\textbf{Model}}
      & \multicolumn{2}{c}{\textbf{Athens}}
      & \multicolumn{2}{c}{\textbf{Lisbon}}
      & \multicolumn{2}{c}{\textbf{Rome}} \\
    \cline{3-8}
      &
      & $\beta_{\text{HWD}_t}$ & $\beta_{\text{CWD}_t}$
      & $\beta_{\text{HWD}_t}$ & $\beta_{\text{CWD}_t}$
      & $\beta_{\text{HWD}_t}$ & $\beta_{\text{CWD}_t}$ \\
    \midrule
    20--64 & DLNM--LC & $0.0132^{***}$ & $0.0067$ & $0.0138^{**}$ & $0.0100^{**}$ & $0.0038$ & $0.0077^{**}$ \\
     & DLNM--LL & $0.0132^{***}$ & $0.0069$ & $0.0140^{**}$ & $0.0104^{***}$ & $0.0037$ & $0.0088^{**}$ \\
    \midrule
    65--74 & DLNM--LC & $0.0095^{***}$ & $0.0097^{**}$ & $0.0205^{***}$ & $0.0039$ & $-0.0112^{**}$ & $0.0015$ \\
     & DLNM--LL & $0.0094^{***}$ & $0.0107^{***}$ & $0.0206^{***}$ & $0.0039$ & $-0.0117^{***}$ & $0.0010$ \\
    \midrule
    75--84 & DLNM--LC & $0.0161^{***}$ & $0.0060^{*}$ & $0.0181^{***}$ & $0.0125^{***}$ & $0.0089^{**}$ & $0.0066^{**}$ \\
     & DLNM--LL & $0.0161^{***}$ & $0.0053$ & $0.0182^{***}$ & $0.0121^{***}$ & $0.0081^{**}$ & $0.0052$ \\
    \midrule
    85+ & DLNM--LC & $0.0198^{***}$ & $0.0007$ & $0.0198^{***}$ & $0.0134^{***}$ & $0.0184^{***}$ & $0.0080^{**}$ \\
     & DLNM--LL & $0.0197^{***}$ & $0.0004$ & $0.0199^{***}$ & $0.0130^{***}$ & $0.0177^{***}$ & $0.0070^{**}$ \\
\bottomrule
\noalign{\vskip 6pt}
\end{tabular}
}
\footnotesize{Level of significance: $0.01$:***, $0.05$:**, $0.1$:*}
\caption{\small Coefficients of $\text{HWD}_t$ and $\text{CWD}_t$ in the DLNM component of DLNM--LC and DLNM--LL models.}
\label{hwd_cwd_coefficients_combined}
\end{table}

\subsection{Climate loadings}
\begin{figure}[H]
\centering
\includegraphics[width=0.91\linewidth]{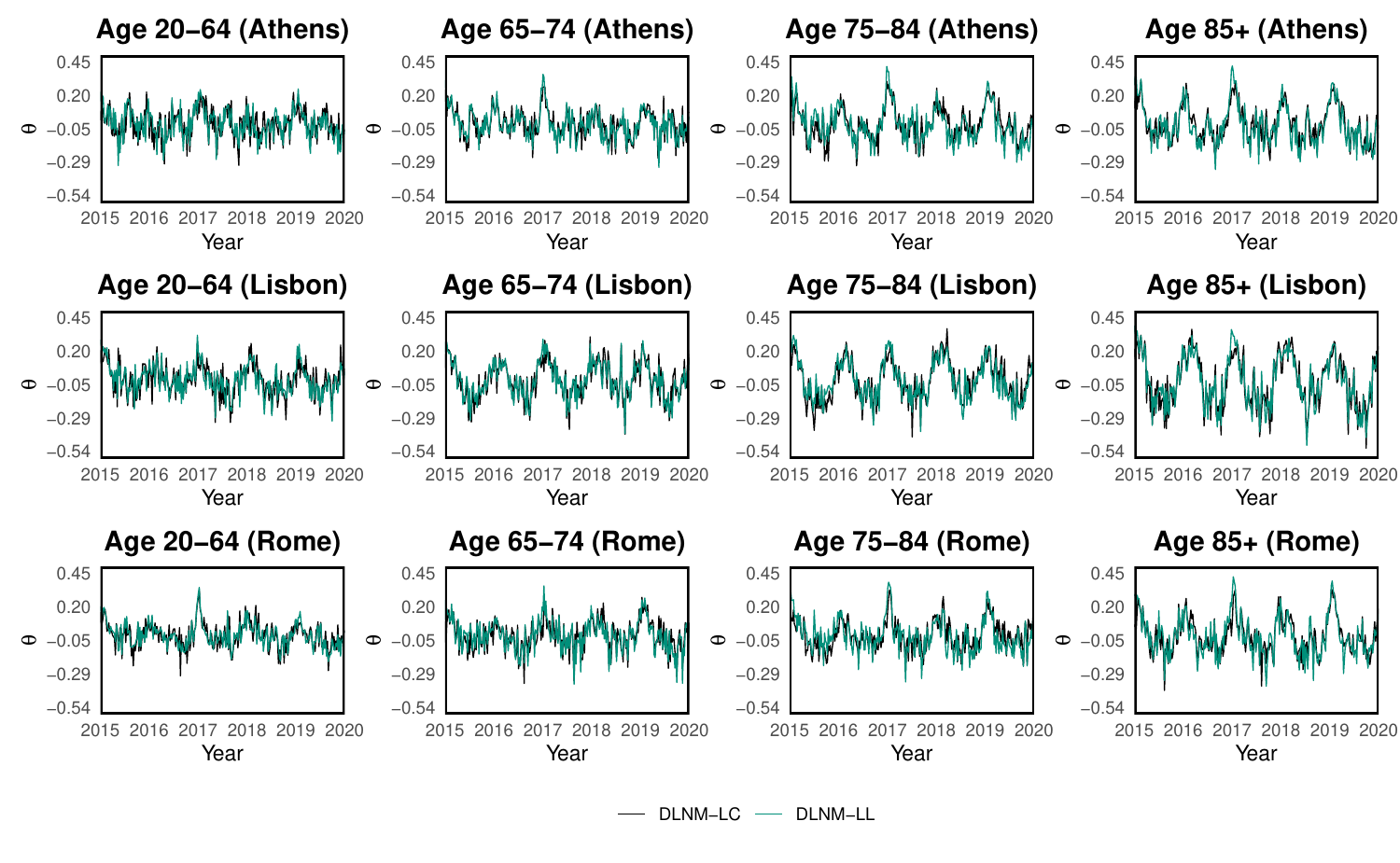}
\vspace{-0.25in}
\caption{Climate loading ${\theta}$ for each age group in the three regions.}
\label{loading}
\end{figure}

Recall that the climate loading $\theta$, defined in Equation (\ref{theta}), measures the proportion of climate-driven mortality relative to the total mortality. It provides an alternative way to assess the strength of seasonal patterns in mortality. In Figure \ref{loading}, we plot the estimated climate loadings from both the DLNM--LC and DLNM--LL models. 

For both models, the climate loading plots exhibit seasonal patterns, with the strength of seasonality increasing with age. Among the four age groups, those aged 20--64 have the smallest range (around -25\%--30\%) in climate loading, while those aged 85+ have the largest (around -35\%--45\%). In winter months, the value of $\theta$ is generally positive, while in summer months,  $\theta$ can be both negative and positive. The positive loading values in summer reflect the adverse impact of extreme heat on mortality. Regional differences in climate loadings can also be seen from the plots, with Lisbon exhibiting a wider range in climate loading, especially during the summer months when mortality compensation is present. For Athens and Rome, notably high loading values are observed during the winter of 2017 for all age groups, attributed to excess mortality following an anomalously cold January in southeastern Europe \citep{demirtacs2022anomalously}. 
\vspace{-0.15in}
\vspace{-0.25in}
\subsection{Expanding window cross-validation}
\label{expanding-window-cv}
In this subsection, we perform a cross-validation exercise using an expanding window approach across six models, including the original Lee--Carter model (referred to as ``LC''), the original Li--Lee model (referred to as ``LL''), the DLNM--LC model,  the DLNM--LL model, the DLNM model by \cite{madaniyazi2024seasonality}  which separately models the non-temperature-driven long-term trend and temperature-driven seasonality\footnote{Please refer to online supplementary material Section G.1 for details.},  and the two-stage DLNM approach by \cite{GUIBERT2025}\footnote{Please refer to online supplementary material Section G.2 for details.}.
We divide the mortality data from 2015 to 2019 into training sets and test sets for expanding window cross-validation. 
We set up a 10-fold expanding window with an initial training size of 102 weeks, an expansion step of 8 weeks, and a forecasting horizon of 78 weeks.\footnote{Note that the stochastic components of the models are predicted using a time series approach, while the UTCI data is considered given and the actual observed data during the validation period is used.} With an initial training window of 102 weeks, we fit the mortality model on the training set and generate forecasts for the subsequent 78 weeks. The forecast accuracy is then evaluated by comparing the forecasts with the observed mortality over this 78-week test window. We then expand the training window by adding the next 8 weeks and repeat the same procedure. This process is repeated 10 times, yielding 10 expanding-window splits.

To evaluate forecasting performance, the Mean Absolute Error (MAE) for the 78-week-ahead forecasts for region $i$ and age group $x$ is defined as follows:
\vspace{-0.05in}
\begin{equation}
\text{MAE}(x,i) = \frac{1}{10 \times 78}\sum_{j=1}^{10} \sum_{t=1}^{78} \left|m(x,102+t+8j,i)-\hat{m}(x,102+t+8j,i)\right|.
\end{equation}
\vspace{-0.08in}

Table \ref{tab:compare_mae} reports MAE values (scaled by a factor of 100) from our cross-validation exercise.  In the table, the best-performing model is highlighted in bold, and the second-best-performing model is highlighted in italics.  
The results demonstrate that the proposed DLNM–LC and DLNM–LL models achieve higher forecasting accuracy across age groups and regions than their corresponding baseline stochastic models. This confirms the importance of incorporating climate effects into stochastic mortality models to achieve more accurate total mortality projections, at least over the forecast horizon considered in this study. Overall, our proposed models achieve better performance in Athens and Rome, whereas \cite{madaniyazi2024seasonality} performs better in Lisbon. It should be noted that \cite{madaniyazi2024seasonality} is a strong benchmark relative to the other models considered. These results thus provide reassurance that our proposed models perform well in short-term forecasting
    while offering a joint modeling framework across age and regions, and enabling the separation of common time trends from region-specific trends. 
Between DLNM--LC and DLNM--LL, we argue that DLNM--LL has slightly stronger performance. 

    \begin{table}[H]
    \centering
    
    \resizebox{\textwidth}{!}{
    \begin{tabular}{ccccccc}
        \toprule
        \small\textbf{Athens} 
          & \small\textbf{LC} 
          & \small\textbf{LL} 
          & \small\textbf{DLNM--LC} 
          & \small\textbf{DLNM--LL} 
          & \small\textbf{\citeauthor{madaniyazi2024seasonality}}
          & \small\textbf{\citeauthor{GUIBERT2025}} \\
        \midrule
        20--64 & 0.0258 & 0.0285 & \textbf{0.0231} & 0.0233 & \textit{0.0232} & 0.0298 \\
        65--74 & 0.1636 & 0.1830 & \textbf{0.1355} & 0.1565 & \textit{0.1399} & 0.1787 \\
        75--84 & 0.5009 & 0.5091 & \textit{0.3675} & 0.3767 & \textbf{0.3551} & 0.4122 \\
        85+    & 1.8120 & 2.0056 & \textbf{1.2836} & 1.5514 & \textit{1.3741} & 1.5648 \\
        \midrule
        \small\textbf{Lisbon} 
          & \small\textbf{LC} 
          & \small\textbf{LL} 
          & \small\textbf{DLNM--LC} 
          & \small\textbf{DLNM--LL} 
          & \small\textbf{\citeauthor{madaniyazi2024seasonality}}
          & \small\textbf{\citeauthor{GUIBERT2025}} \\
        \midrule
        20--64 & 0.0274 & 0.0274 & \textit{0.0257} & \textit{0.0257} & \textbf{0.0244} & 0.0284 \\
        65--74 & 0.1500 & 0.1624 & 0.1592 & \textbf{0.1382} & 0.1527 & \textit{0.1404} \\
        75--84 & 0.4615 & 0.5012 & 0.5032 & \textit{0.4055} & \textbf{0.3850} & 0.5085 \\
        85+    & 1.8087 & 2.0249 & 1.7293 & \textit{1.6425} & \textbf{1.4678} & 1.7378 \\
        \midrule
        \small\textbf{Rome} 
          & \small\textbf{LC} 
          & \small\textbf{LL} 
          & \small\textbf{DLNM--LC} 
          & \small\textbf{DLNM--LL} 
          & \small\textbf{\citeauthor{madaniyazi2024seasonality}}
          & \small\textbf{\citeauthor{GUIBERT2025}} \\
        \midrule
        20--64 & 0.0197 & \textbf{0.0192} & 0.0204 & 0.0206 & \textit{0.0195} & 0.0235 \\
        65--74 & \textbf{0.1189} & \textit{0.1276} & 0.1491 & 0.1475 & 0.1389 & 0.1394 \\
        75--84 & 0.3455 & 0.3927 & 0.3511 & \textbf{0.3254} & \textit{0.3298} & 0.3635 \\
        85+    & 1.4920 & 1.7384 & 1.4388 & \textbf{1.3179} & \textit{1.4258} & 1.7075 \\
        \bottomrule
    \end{tabular}}
    \caption{MAE ($\times 100$) for 10--fold expanding window cross-validation.}
    \label{tab:compare_mae}
\end{table}

It should be noted that the relatively small size of the dataset is a limitation of our empirical study. We therefore acknowledge that expanding-window cross-validation can assess short- to medium-term forecasting performance but has limited relevance for long-term projections.
\vspace{-0.1in}

\section{Mortality projection under RCP scenarios}
\label{sec:Mortality projection under RCP scenarios}
\subsection{Future UTCI data under RCP scenarios}
Representative Concentration Pathways (RCPs) are climate change scenarios that project future greenhouse gas concentrations and their associated radiative forcing levels through to the year 2100 \citep{van2011representative}.  Each RCP represents a different trajectory of radiative forcing, measured in watts per square meter ($\text{W/m}^2$), resulting from different levels of anthropogenic emissions. 
The original framework developed by the Intergovernmental Panel on Climate Change (IPCC) includes four RCP scenarios: RCP2.6, RCP4.5, RCP6.0 and RCP8.5.\footnote{For more information, please see \url{https://en.escarus.com/climate-change-risk-projections-rcp-scenarios}.}

Among the four RCP scenarios, we consider RCP2.6 and RCP8.5 for \textcolor{black}{total} mortality projection, as they represent two extreme cases. 
We compute future UTCI data under the two RCP scenarios by Equation (\ref{utci_eq}).
The input variables, air temperature ($T_a$), relative humidity (RH), and wind speed (WS), are obtained from the Coupled Model Intercomparison Project Phase 5 (CMIP5) climate model.\footnote{Data source: \url{cds.climate.copernicus.eu/datasets/projections-cmip5-daily-single-levels?tab=download}. In CMIP 5 model, we select ensemble member r1i1p1 (initial condition) with two sub-climate models: CSIRO-MK3-6-O (Australia) and BCC-CSM-1-M (China).} 
Across the three variables, the common time period for future projections spans from 2031 to 2045, defining the feasible simulation period.
Since projections for regional-level mean radiant temperature (MRT) are not available, we follow the method of \cite{di2020thermal} and utilize historical MRT data from the ERA5-HEAT dataset to simulate future UTCI values. 
\vspace{-0.1in}
\subsection{Simulation design}
The scenario-based \textcolor{black}{total} mortality projections consist of two main components, designed to incorporate uncertainty from both the stochastic and DLNM components into the forecasts. The sample size for the simulation is 10,000. We first simulate from the stochastic component to generate distinct time series trajectories for the time-varying mortality factors without explicit assumptions about changes in age demographics. 
For the DLNM component, as described in Section \ref{sec:2.1}, we use a bootstrap method to generate the paths of climate-driven mortality components.
Finally, we simulate an independent and identically distributed (\textit{i.i.d.}) Gaussian error term and add it to the projections.
This integrated approach enables us to jointly account for uncertainty in both the inherent long-term mortality trend and the climate-driven mortality forecasts under scenario-based projections. Given the data limitations of the research, the scenario-based projections are intended for application and demonstration of the proposed method only. As such, the conclusions from these projections are subject to the limitations of the available data and the modeling assumptions. 
\vspace{-0.1in}
\subsection{Weekly mortality projections: 2031--2045}
We conduct weekly mortality projections under scenarios RCP2.6 and RCP8.5 for the period 2031--2045. 
For illustrative purposes, in this paper, we only present mortality projections from the DLNM--LL model.\footnote{Consistent findings were obtained from the DLNM--LC model, and additional results are available upon request.}  
Compared to the DLNM--LC model, the DLNM--LL model is better suited for long-term multi-population mortality forecasting due to its incorporation of a common trend, which ensures coherence in mortality forecasts across similar regions without significant divergence. 

\begin{figure}[h!]
\centering
\includegraphics[width=0.72\linewidth]{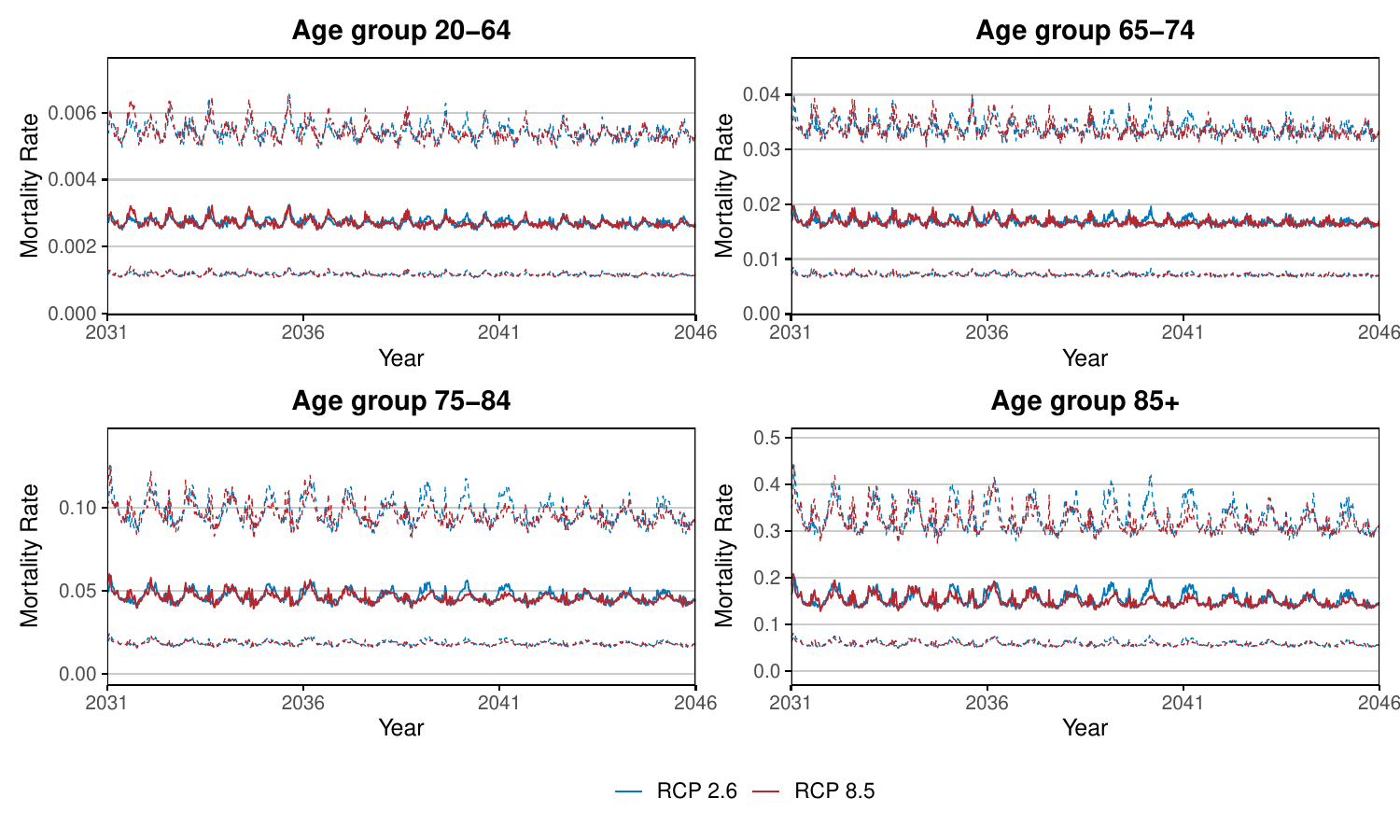}
\vspace{-0.2in}
\caption{Weekly mortality projection for Athens (2031--2045).}
\label{sim.week.attica}
\end{figure}

\begin{figure}[h!]
\centering
\includegraphics[width=0.72\linewidth]{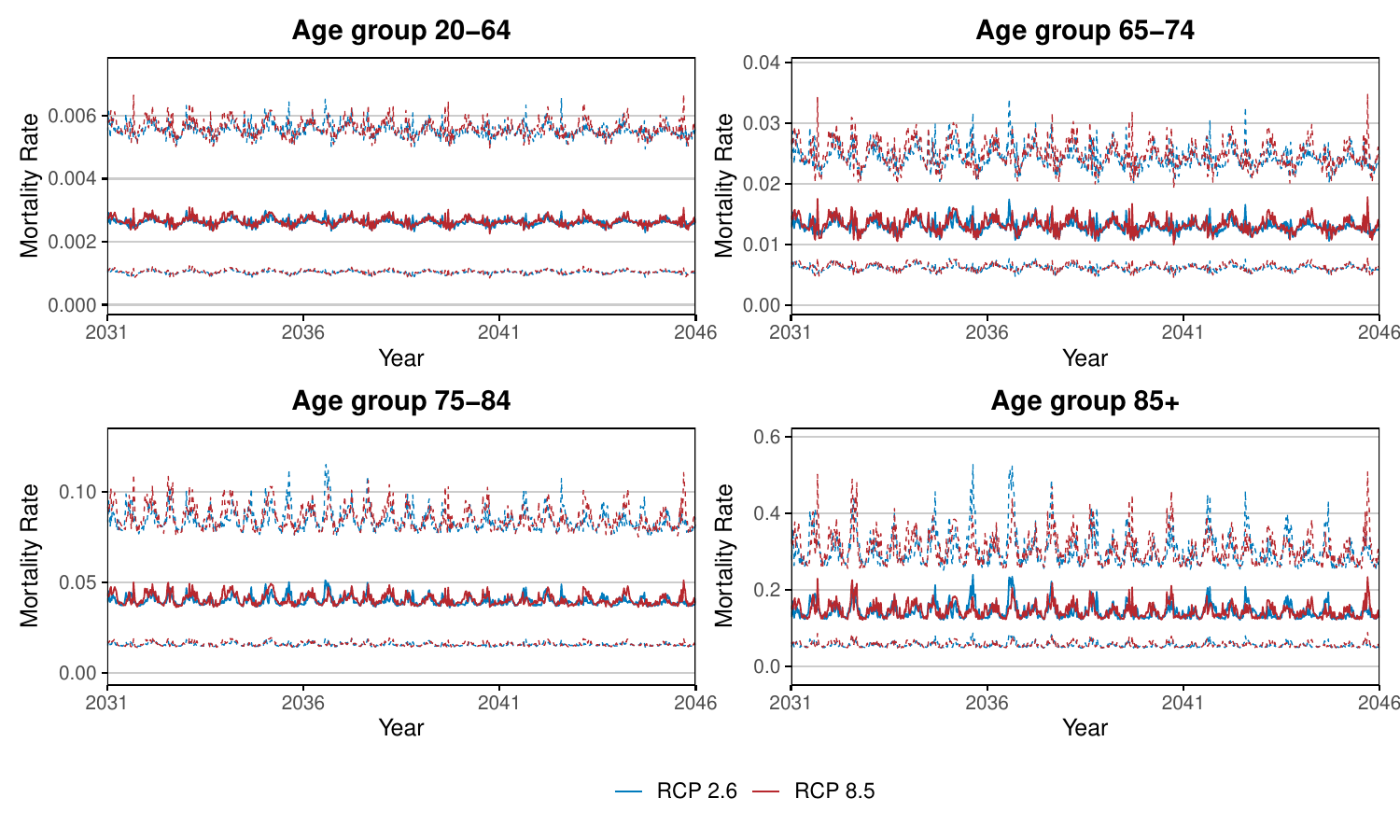}
\vspace{-0.2in}
\caption{Weekly mortality projection for Lisbon (2031--2045).}
\label{sim.week.lisbon}
\end{figure}

\begin{figure}[H]
\centering
\includegraphics[width=0.72\linewidth]{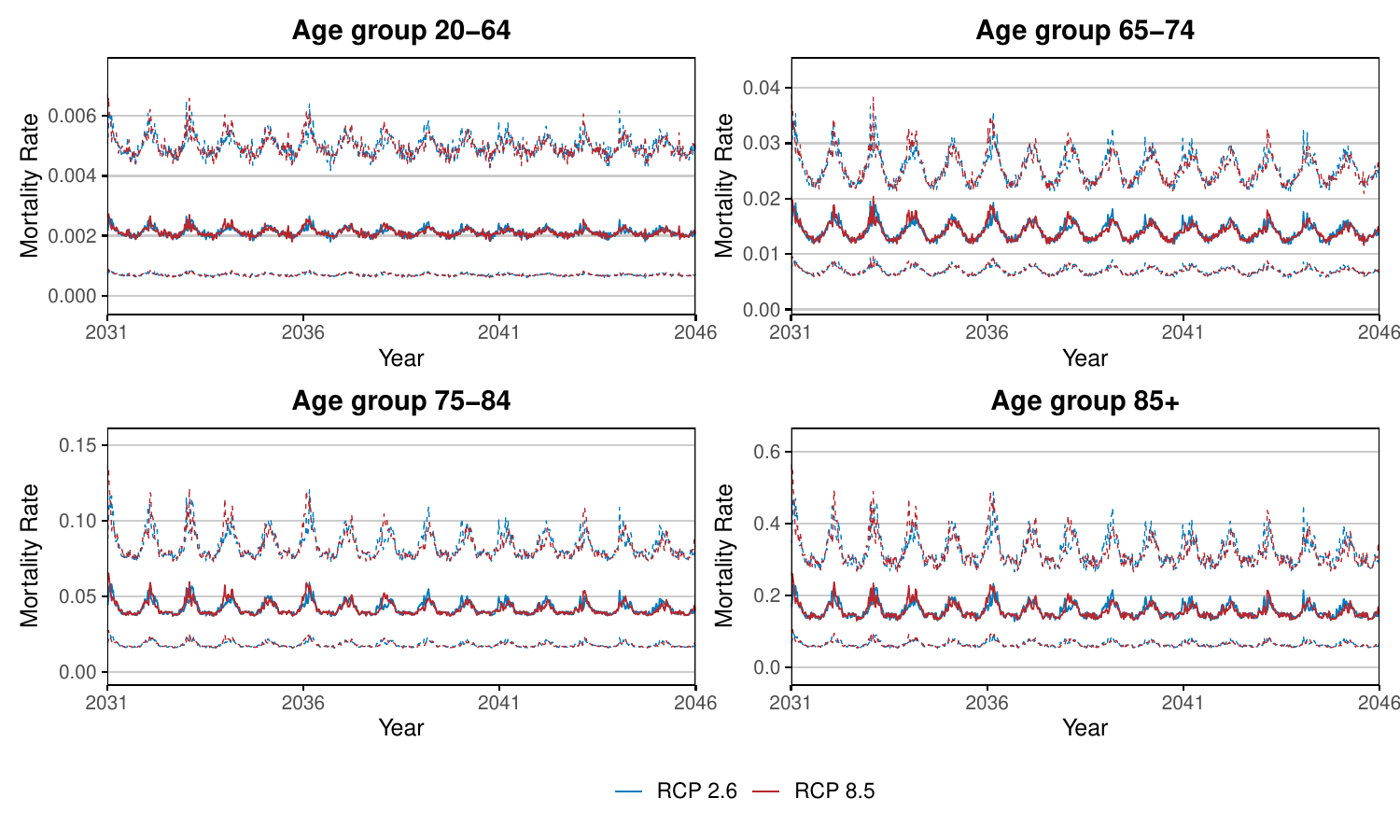}
\vspace{-0.2in}
\caption{Weekly mortality projection for Rome (2031--2045).}
\label{sim.week.rome}
\end{figure}

Weekly mortality projections for Athens, Lisbon, and Rome are shown in Figures \ref{sim.week.attica}, \ref{sim.week.lisbon}, and \ref{sim.week.rome}, respectively. 
The lower and upper dashed lines represent the 2.5th and 97.5th percentiles of the probabilistic forecasts, and the solid line represents the mean of the simulations. This format is consistently applied in all subsequent simulation plots.

Across the three regions, the mortality projections for older age groups 
75--84 and 85+ exhibit stronger seasonality compared to younger age groups, since these cohorts are more vulnerable to both extreme cold and hot weather conditions.
We can see that the seasonal pattern is W-shaped, with the winter mortality peak remaining higher than the summer mortality peak, which is consistent with historical mortality rates. A slight trend of mortality improvement may be observed in some of the plots, though it is difficult to confirm due to the presence of strong seasonality.

Comparing the mortality projections between the RCP2.6 and RCP8.5 scenarios, we observe a lower winter mortality peak under RCP8.5, reflecting the compensating effect of warmer weather conditions on winter mortality. This will lead to a reduction in cole-related mortality. As a result, we expect winter mortality rates to be generally lower under RCP8.5 than under RCP2.6.
\vspace{-0.1in}
\vspace{-0.15in}

On the other hand, the higher UTCI under RCP8.5 will intensify extreme hot weather conditions in summer, thereby increasing heat-related mortality compared to RCP2.6.
As a result, we expect summer mortality rates to be relatively higher under RCP8.5 than under RCP2.6.
While these projections clearly highlight contrasting seasonal effects of climate change, the overall impact on annual mortality remains less straightforward. This will be examined in the following section.
\vspace{-0.1in}

\subsection{Annual mortality projections: 2031--2045}

In this section, we present annual mortality projections for the same time period across the three regions\footnote{Since we assume uniform risk exposure throughout the year, annual mortality is calculated as the summation of weekly mortality divided by 52.}.
By examining annualized mortality rates, we smooth out short-term fluctuations within each year and capture the cumulative effects of climate on mortality. In other words, we look at the combined impact of winter mortality compensation and summer excess mortality under RCP2.6 and RCP8.5.
To ensure consistency, we present the results from the DLNM--LL model\footnote{Results based on the DLNM--LC model are available upon request.}.

\begin{figure}[H]
\centering
\includegraphics[width=0.72\linewidth]{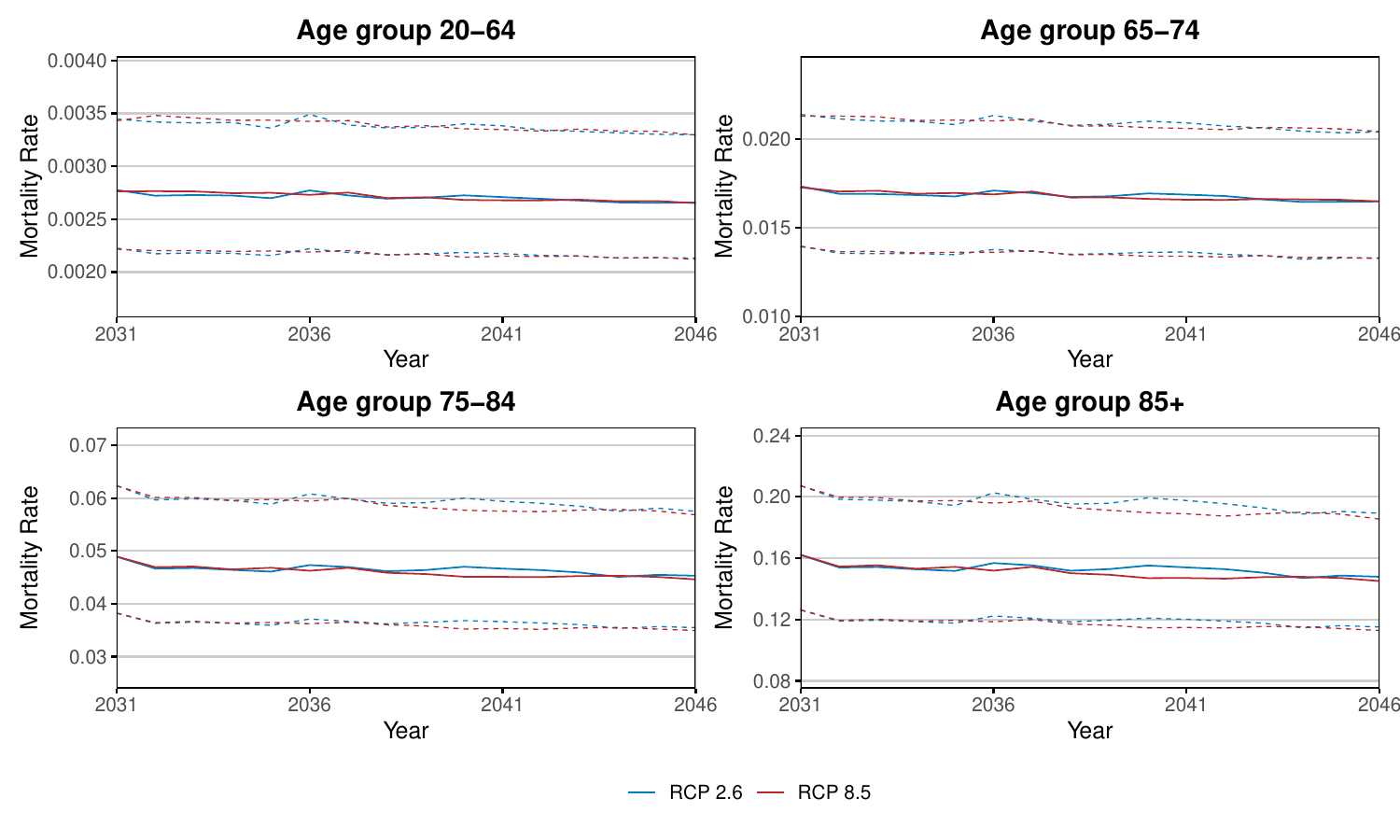}
\vspace{-0.2in}
\caption{Annualized mortality projection for Athens (2031--2045).}
\label{sim.year.attica}
\vspace{-0.15in}
\end{figure}

For illustrative purposes, we plot the annualized mortality projections for Athens shown in Table \ref{sim.year.attica}.\footnote{Annualized mortality projections for Lisbon and Rome are shown in online supplementary material Section F.}
Over the 15-year projection period, we observe a general downward trend in mortality, primarily driven by improvements in the stochastic mortality components. 
When comparing the two RCP scenarios, we observe a lower level of mortality under RCP8.5 in Athens, particularly from the mid 2030s to the early 2040s. This suggests that the impact of rising UTCI on winter mortality compensation outweighs its effect on excess summer mortality. The mortality reduction is especially pronounced in old age groups 75--84 and 85+ compared to those aged 20--64 and 65--74. However, toward the end of the forecasting period, we observe a tendency for the mortality rate to be higher under RCP8.5 compared to RCP2.6, indicating a reversal in the relative influence of the two opposing effects. 
Overall, we observe that the annualized mortality rates for ages 20--64 and 65--74 are not substantially different under the two RCP scenarios compared to the other two age groups.

Our results are consistent with those findings in \cite{lee2023future}, which provide additional reassurance on the correctness of our integration of DLNM into stochastic mortality modeling. By explicitly accounting for long-term stochastic mortality risk, our modeling framework allows us to quantify and evaluate the impact of climate change on forecasts of seasonal mortality patterns and overall mortality trends. In the short run, reductions in winter mortality may offset increases in summer mortality, accelerating mortality improvement rates. However, in the long run, the rise in summer excess mortality is expected to catch up and eventually outweigh the winter reductions, leading to a slowdown in mortality improvement or even an increase in total mortality over time. This dynamic suggests a shift in the seasonal burden of mortality and predicts a potential long-term increase in total mortality under the RCP8.5 scenario. Importantly, the implications arising from the combined effect of long-term stochastic mortality trend and climate-driven mortality have not been adequately discussed in other similar studies. 
\vspace{-0.1in}

\section{Conclusion}
\label{sec:Conclusion}
This paper introduces novel single- and multi-population mortality forecasting models, the DLNM--LC model and DLNM--LL model, which integrate a DLNM into stochastic models to forecast long-term mortality levels under different climate scenarios.
These models are applied to weekly mortality data from Athens, Lisbon, and Rome for the period 2015--2019. We compare the estimated time-varying factors in the stochastic mortality components with those from the Lee--Carter and Li--Lee models. Next, we examine the climate-driven mortality components using DLNM, which reveals a clear U-shaped relationship between climate-related mortality risk and UTCI. 
Furthermore, we implement an expanding-window cross-validation exercise and find that both proposed models provide strong forecasting performance over short-term forecasting horizons. 
Finally, based on the fitted models, we conduct scenario-based mortality forecasts for the period 2031--2045 under RCP2.6 and RCP8.5. The results suggest that overall mortality could initially decrease under RCP8.5 due to warmer winter conditions, but may increase as extreme heat events become more intense over time.

The proposed modeling framework can be easily adapted to different applications by adjusting the stochastic mortality components and the DLNM components.
Future research could explore pooling data across finer age groups and more geographical locations to improve estimation accuracy, although this would depend on the availability of such detailed data.  The long-term mortality projections in this research are included as an illustrative application of the proposed method and should not be taken as more than very tentative evidence for future mortality, given the short data sets and that evaluation is only conducted for short-term forecasts. When longer data series become publicly available, future studies can provide a more thorough validation of long-term projections.

\bibliography{reference1}
\clearpage
\begin{center}
    {\LARGE \textbf{Supplementary Material}}
\end{center}
\appendix
\counterwithin{figure}{section}
\section{Model algorithm and estimation}
\subsection{Lee--Carter model (SVD method)}
\begin{algorithm}
\caption{Estimation of Lee--Carter Model (SVD Method)}
\setstretch{1.3}
\KwInput{an $N \times T$ (age $\times$ time) mortality matrix $\bm{m} = \left[m(x,1), \dots,m(x,T) \right]$}
\KwOutput{$\log \hat{\bm{m}}$ for $t = 1,2, \dots, T$, $\hat{\bm{\kappa}} = \left[\hat{\kappa}(1), \hat{\kappa}(2), \dots, \hat{\kappa}(T)\right]$, $\hat{\bm{a}} = \left[\hat{a}(1), \hat{a}(2), \dots, \hat{a}(N)\right]'$, and $\hat{\bm{b}} = \left[\hat{b}(1), \hat{b}(2), \dots, \hat{b}(N)\right]'$.}
\KwProcedure{LC($\bm{m}$)}
\begin{algorithmic}[1]
\setstretch{1.5}
\State $\hat{a}(x) \leftarrow \frac{1}{T}\sum_{t=1}^T \log m(x,t)$, $\forall x$.
\State Combine $\log \bar{m}(x, t) \leftarrow \log m(x,t) - \hat{a}(x)$, $\forall t$ as a matrix form $\bar{\bm{m}} = \left[\bar{m}(x,1),\dots, \bar{m}(x,T)\right]$.
\State Perform SVD on $\log \bar{\bm{m}}$:
$$\log \bar{\bm{m}} = \bm{U} \bm{D} \bm{V'}$$ where $\bm{U}$ and $\bm{V}$ are left singular vector and right singular vector respectively.
\State $\hat{\bm{\kappa}} \leftarrow d_1 \bm{v} \sum_{i = 1}^{N} u_i$, where $d_1$ is the first singular value, $\bm{u}$ and $\bm{v}$ are the first left and right singular vector of $\bar{\bm{m}}$, respectively.
\State $\hat{\bm{b}} \leftarrow \bm{u} / \sum_{i = 1}^{N} u_i, \forall x$.
\State $\log \hat{m}(x, t) \leftarrow \hat{a}(x) + \hat{b}(x)\hat{\kappa}(t),  \forall x, t$.
\end{algorithmic}
\label{LC}
\end{algorithm} 
\subsection{Li--Lee model (Product-ratio method)}
\begin{algorithm}[H]
\label{generalizedLL}
\caption{Estimation of the Lee--Li Model (Product-Ratio Method)}
\setstretch{1.5}
\KwInput{an $N \times T \times n$ (age $\times$ time $\times$ population index) mortality tensor $\bm{M} = [\bm{m}(x,t,1), \dots, \bm{m}(x,t,n)]$}
\KwOutput{$\hat{\bm{A}} = [\hat{\bm{A}}(x,1), \dots, \hat{\bm{A}}(x,n)]$, $\hat{\bm{B}}$, $\hat{\bm{K}}$, $\hat{\bm{b}} = [\hat{\bm{b}}(x,1), \dots, \hat{\bm{b}}(x,n)]$, $\hat{\bm{\kappa}} = [\hat{\bm{\kappa}}(t,1), \dots, \hat{\bm{\kappa}}(t,n)]$, and $\hat{\bm{M}} = [\hat{\bm{m}}(x,t,1), \dots, \hat{\bm{m}}(x,t,n)]$}
\KwProcedure{LL($\bm{M}$)}
$p(x,t) \leftarrow \left(\prod_{j=1}^n m(x,t,j)\right)^{1/n}, \forall x, t$
$\hat{\bm{A}}_p, \hat{\bm{B}}, \hat{\bm{K}}, \hat{\bm{m}}_p \leftarrow LC(p(x,t)), \forall x, t$.
\For{$1 \leq i \leq n$}
{
$r(x,t,i) \leftarrow m(x,t,i)/p(x,t)$
$\hat{\bm{a}}, \hat{\bm{b}}, \hat{\bm{\kappa}}, \hat{\bm{m}}_r \leftarrow LC(r(x,t,i))$
$\hat{\bm{A}} \leftarrow \hat{\bm{A}}_p + \hat{\bm{a}}$
$\log \hat{\bm{m}} \leftarrow \log \hat{\bm{m}}_p + \log \hat{\bm{m}}_r$
}
$\hat{\bm{M}} \leftarrow [\hat{\bm{m}}(x,t,1), \dots, \hat{\bm{m}}(x,t,n)], \forall x, t$
$\hat{\bm{A}} \leftarrow [\hat{\bm{A}}(x,1), \dots, \hat{\bm{A}}(x,n)], \forall x$
$\hat{\bm{b}} = [\hat{\bm{b}}(x,1), \dots, \hat{\bm{b}}(x,n)], \forall x$
$\hat{\bm{\kappa}} = [\hat{\bm{\kappa}}(t,1), \dots, \hat{\bm{\kappa}}(t,n)]$, $\forall t$
\end{algorithm}

\subsection{The algorithm of DLNM-LL model}
In this section, we provide detailed DLNM-LL estimation.
The estimation procedure is described as follows: First, we fit the Li--Lee model on $\log \hat{\bm{M}}^{(0)}$ to obtain the parameter set $\bm{\Theta}^{(0)}_{\text{LL}} = \left\{\hat{\bm{A}}^{(0)}, \hat{\bm{B}}^{(0)}, \hat{\bm{K}}^{(0)}, \hat{\bm{a}}^{(0)}, \hat{\bm{b}}^{(0)}, \hat{\bm{\kappa}}^{(0)}\right\}$ in the first recursion, and remove the age-specific mean $\hat{\bm{A}}^{(0)} = \left[\hat{\bm{A}}^{(0)}(x,1), \dots, \hat{\bm{A}}^{(0)}(x,n) \right]$ from the log mortality tensor $\log \hat{\bm{M}}^{(0)}$ by each region $i = 1, \dots, n$. We define the remaining component as the partial residual tensor, which is denoted as $\hat{\bm{e}}^{(0)} = \left[\hat{\bm{e}}^{(0)}_1, \dots, \hat{\bm{e}}^{(0)}_n \right]$, where $\hat{\bm{e}}^{(0)}_i = \left[\hat{\bm{e}}^{(0)}_{1,i}, \dots, \hat{\bm{e}}^{(0)}_{N,i} \right]$. 
We then fit the DLNM to partial residuals, separately for each age group $x$ and each region $i$:
\begin{equation}
    \hat{\bm{e}}^{(0)}_{x,i} = \mathcal{S}_{x,i}^{(0)}\left(U_{\tau_{t,i}}, \dots, U_{\tau_{t,i}-L}, \text{HWD}_{t,i}, \text{CWD}_{t,i}\right),
\end{equation}
for $x = 1,\dots, N$ and $i = 1, \dots, n$. After that, the fitted DLNM components are removed from the log mortality tensor:
\begin{equation}
    \log \hat{\bm{M}}^{(1)} = \log \hat{\bm{M}}^{(0)} - \hat{\bm{e}}^{(0)},
\end{equation}
where $\log \hat{\bm{M}}^{(1)}$ is the new log mortality matrix for next recursion. We then fit the Li--Lee model on $\log \hat{\bm{M}}^{(1)}$ and obtain the partial residual tensor $\hat{\bm{e}}^{(1)} = \left[\hat{\bm{e}}^{(1)}_{1}, \dots, \hat{\bm{e}}^{(1)}_{n} \right]$. The partial residuals $\hat{\bm{e}}^{(1)}$ are fitted by the DLNM separately by each age group and each region. By repeating these steps for $\log \hat{\bm{M}}^{(j)}$ and $\hat{\bm{e}}^{(j)}$ with $j \geq 1$, we obtain the parameter set in the $j^{th}$ recursion $\bm{\Theta}_{\text{LL}}^{(j)}$, and the $(j+1)^{th}$ log mortality tensor:
\begin{equation}
    \log \hat{\bm{M}}^{(j+1)} = \log \hat{\bm{M}}^{(j)} - \hat{\bm{e}}^{(j)}.
\end{equation}
Similar to the DLNM--LC model, the algorithm stops by either meeting a converging condition
\begin{equation}
\sup_{\bm\theta\in\bm{\Theta}_{\text{LL}}}\Vert\bm\theta^{(j)} - \bm\theta^{(j-1)}\Vert_\infty < \delta,
\end{equation}
or reaching the maximum number of iterations $J$. 
This algorithm is summarized in Algorithm 2. 
\newpage
\section{Equivalence of summed DLNMs and single DLNM} 
\label{sec:Equivalence of multiple DLNMs and single DLNM}
We rewrite the fitted mortality rates under log scale in general form:
\begin{equation}
\label{fitted_model}
\log \hat{m}(x,t,i) = \log \tilde{m}(x,t,i) +  \mathcal{S}_{x,i}\left(U_{\tau_{t,i}}, \dots, U_{\tau_{t,i}-L}, \text{HWD}_{t,i}, \text{CWD}_{t,i}\right),
\end{equation}
where $\log \tilde{m}(x,t,i)$ represents the stochastic mortality component. Specifically, if it is obtained from single-population mortality model, region index $i$ can be ignored. According to DLNM--LC and DLNM--LL model, we have the following recursive expression:
\begin{equation}
\log \hat{m}^{(j)}(x,t,i) = \log \tilde{m}^{(j-1)}(x,t,i) +  \mathcal{S}_{x,i}^{(j-1)}\left(U_{\tau_{t,i}}, \dots, U_{\tau_{t,i}-L}, \text{HWD}_{t,i}, \text{CWD}_{t,i}\right),
\end{equation}
with initial condition $\log \hat{m}^{(0)}(x,t,i) = \log m(x,t,i)$. 

If the algorithm converges or stops in the $r^{\text{th}}$ recursion with $1 \leq r \leq J$, we have the explicit expression of fitted values:
\begin{equation}
\label{summed_fitted_model}
\log \hat{m}(x,t,i) = \log \tilde{m}(x,t,i) + \sum_{j=0}^{r-1} \mathcal{S}_{x,i}^{(j)}\left(U_{\tau_{t,i}}, \dots, U_{\tau_{t-L,i}}, \text{HWD}_{t,i}, \text{CWD}_{t,i}\right)
\end{equation}
Note that each $\mathcal S_{x, i}^{(j)}$ can be represented as $\bm{X'} \bm\zeta^{(j)}$, where $\bm X$ is a fixed design matrix shared across all iterations. Therefore, the sum of smooth terms
\begin{equation}
\sum_{j=0}^{r-1} \mathcal{S}_{x,i}^{(j)}\left(U_{\tau_{t,i}}, \dots, U_{\tau_{t-L,i}}, \text{HWD}_{t,i}, \text{CWD}_{t,i}\right) = \sum_{j=0}^{r-1} \bm{X'}   \bm{\zeta}^{(j)},
\end{equation}
is equivalent to a single smooth term  
\begin{equation}
 \mathcal{S}_{x,i}\left(U_{\tau_{t,i}}, \dots, U_{\tau_{t-L,i}}, \text{HWD}_{t,i}, \text{CWD}_{t,i}\right) = \bm{X'} \bm \zeta,
\end{equation}
with coefficients
\begin{equation}
    \bm{\zeta} := \left(\bm{\beta}, \bm{\eta}\right) = \sum_{j=0}^{r-1}\bm{\zeta}^{(j)}.
\end{equation}
Thus, applying multiple DLNM iterations within the backfitting algorithm is mathematically equivalent to fitting single DLNM on the climate-driven mortality components. The summed coefficients from multiple DLNMs are equal to the coefficients from the single DLNM. 

\newpage
\section{Figures of specific lagged effects via DLNM}
\label{sec:lagged_effects}
\begin{figure}[H]
  \centering
  \includegraphics[width=1\linewidth]{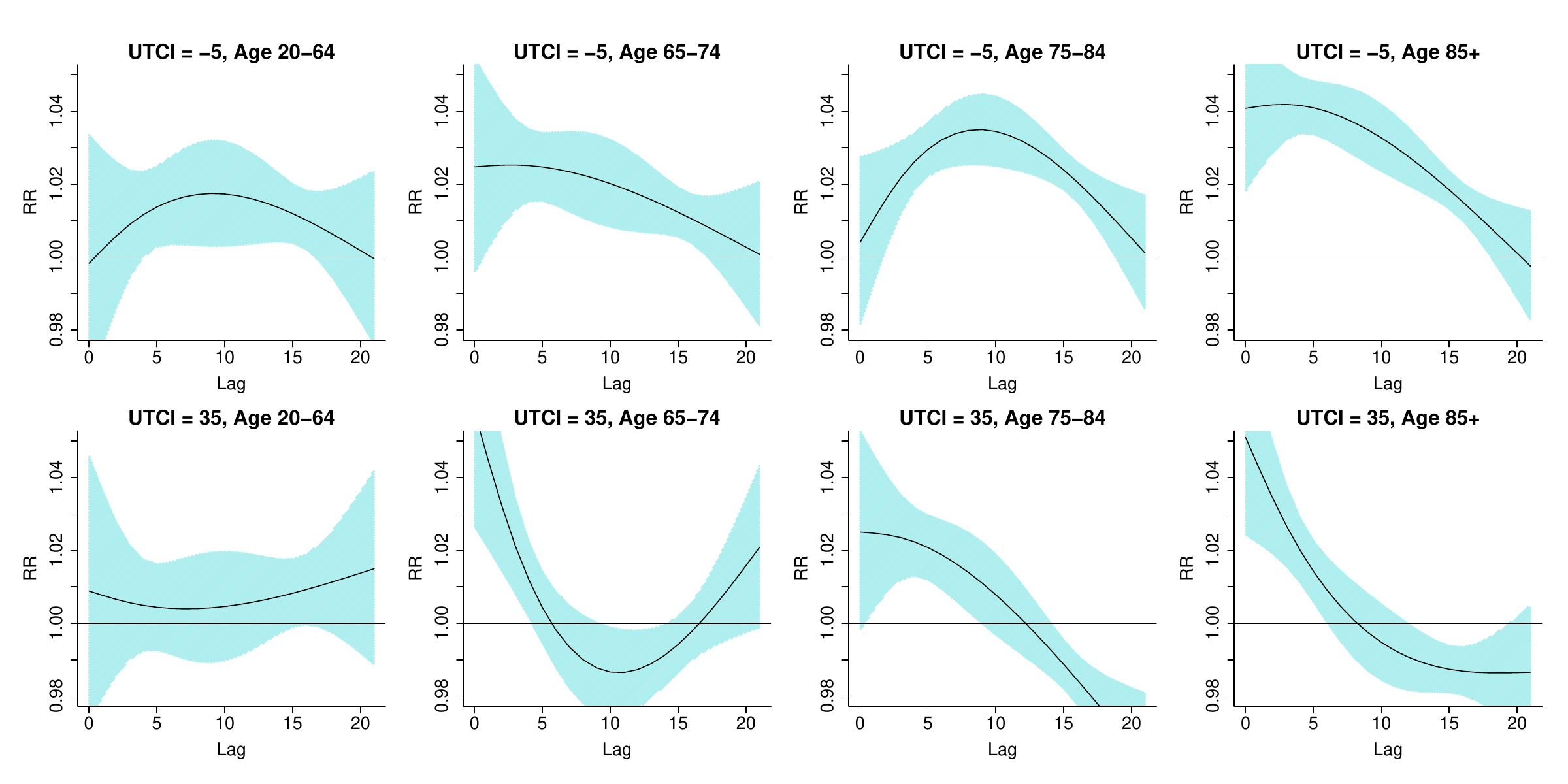}
  \caption{Lagged effect of UTCI = $-5$ and 35 in DLNM--LC model for Athens.}
  \label{fig:lag_athens_lc}
\end{figure}
\begin{figure}[H]
  \centering
  \includegraphics[width=1\linewidth]{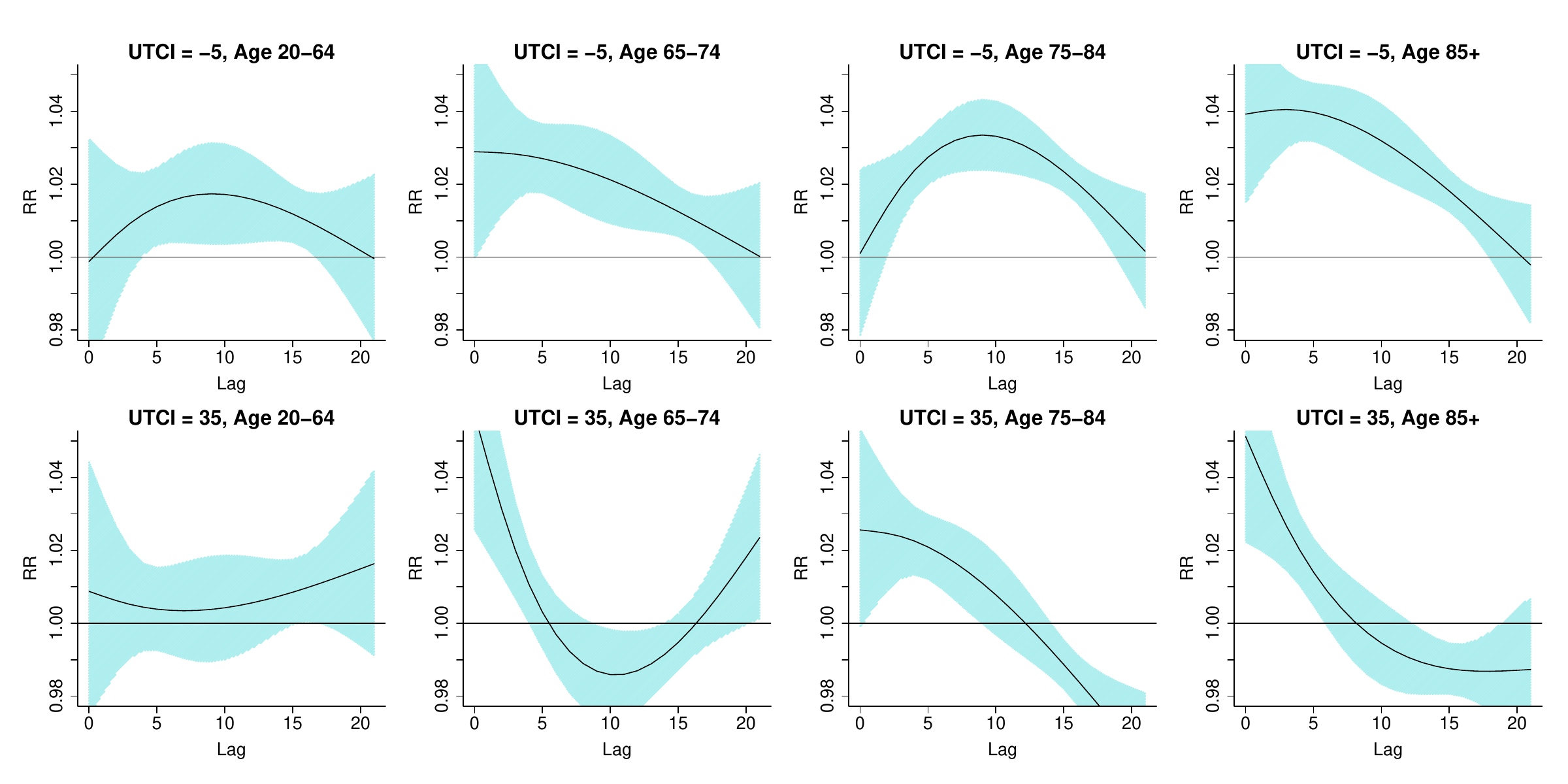}
  \caption{Lagged effect of UTCI = $-5$ and 35 in DLNM--LL model for Athens.}
  \label{fig:lag_athens_ll}
\end{figure}
\begin{figure}[H]
  \centering
  \includegraphics[width=1\linewidth]{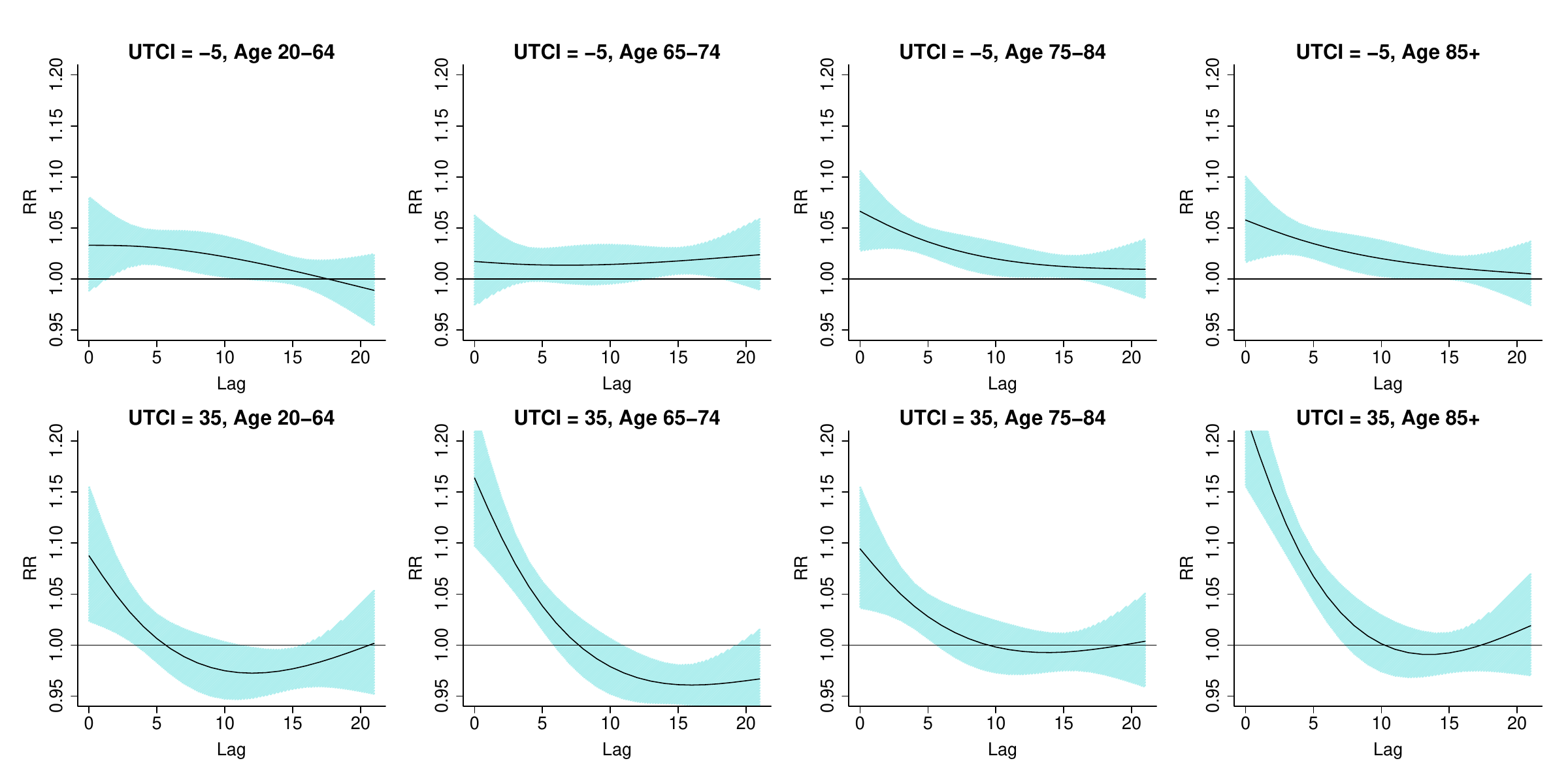}
  \caption{Lagged effect of UTCI = $-5$ and 35 in DLNM--LC model for Lisbon.}
  \label{fig:lag_lisbon_lc}
\end{figure}
\begin{figure}[H]
  \centering
  \includegraphics[width=1\linewidth]{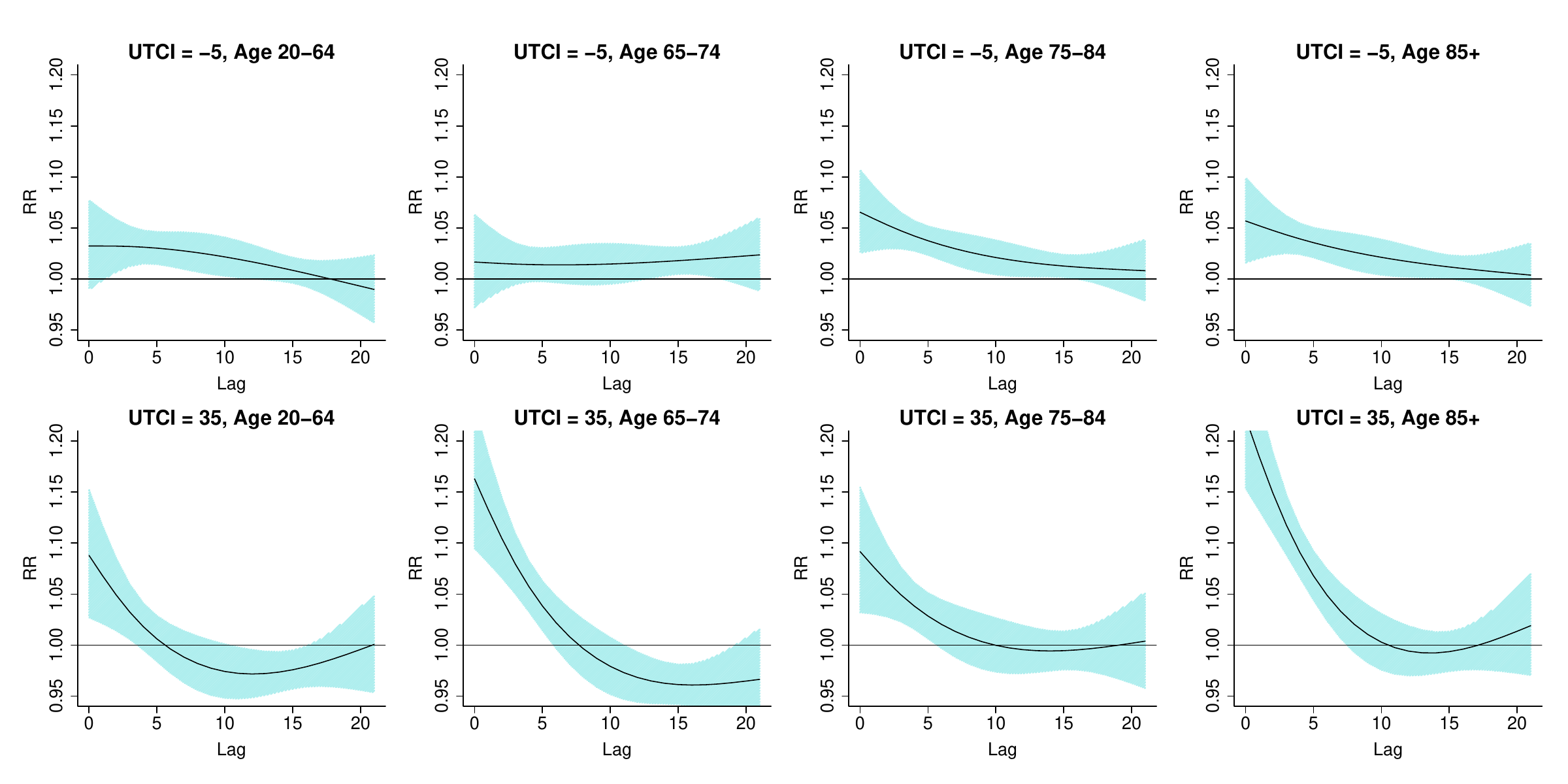}
  \caption{Lagged effect of UTCI = $-5$ and 35 in DLNM--LL model for Lisbon.}
  \label{fig:lag_lisbon_ll}
\end{figure}
\begin{figure}[H]
  \centering
\includegraphics[width=1\linewidth]{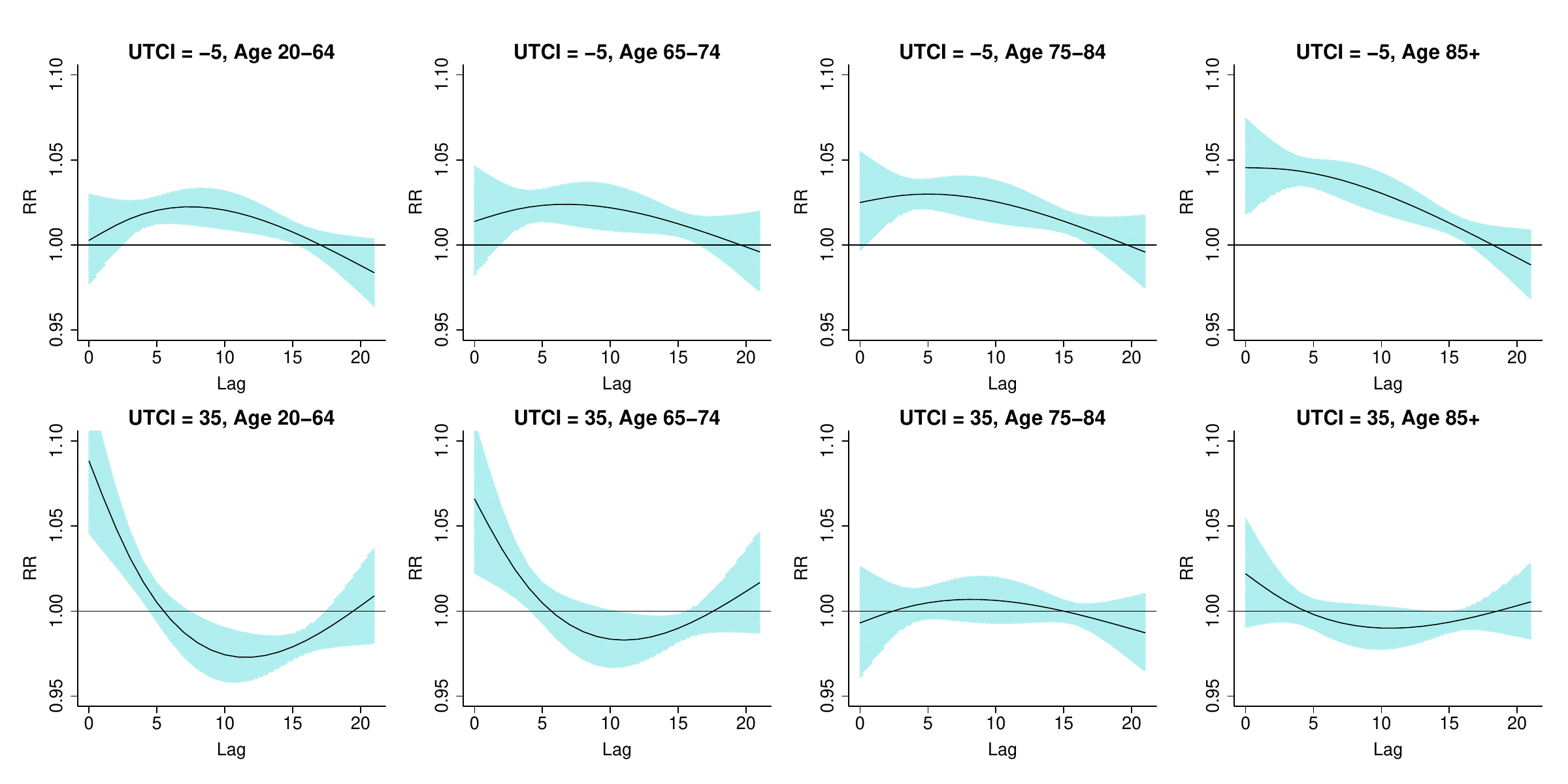}
  \caption{Lagged effect of UTCI = $-5$ and 35 in DLNM--LC model for Rome.}
  \label{fig:lag_rome_lc}
\end{figure}
\begin{figure}[H]
  \centering
  \includegraphics[width=1\linewidth]{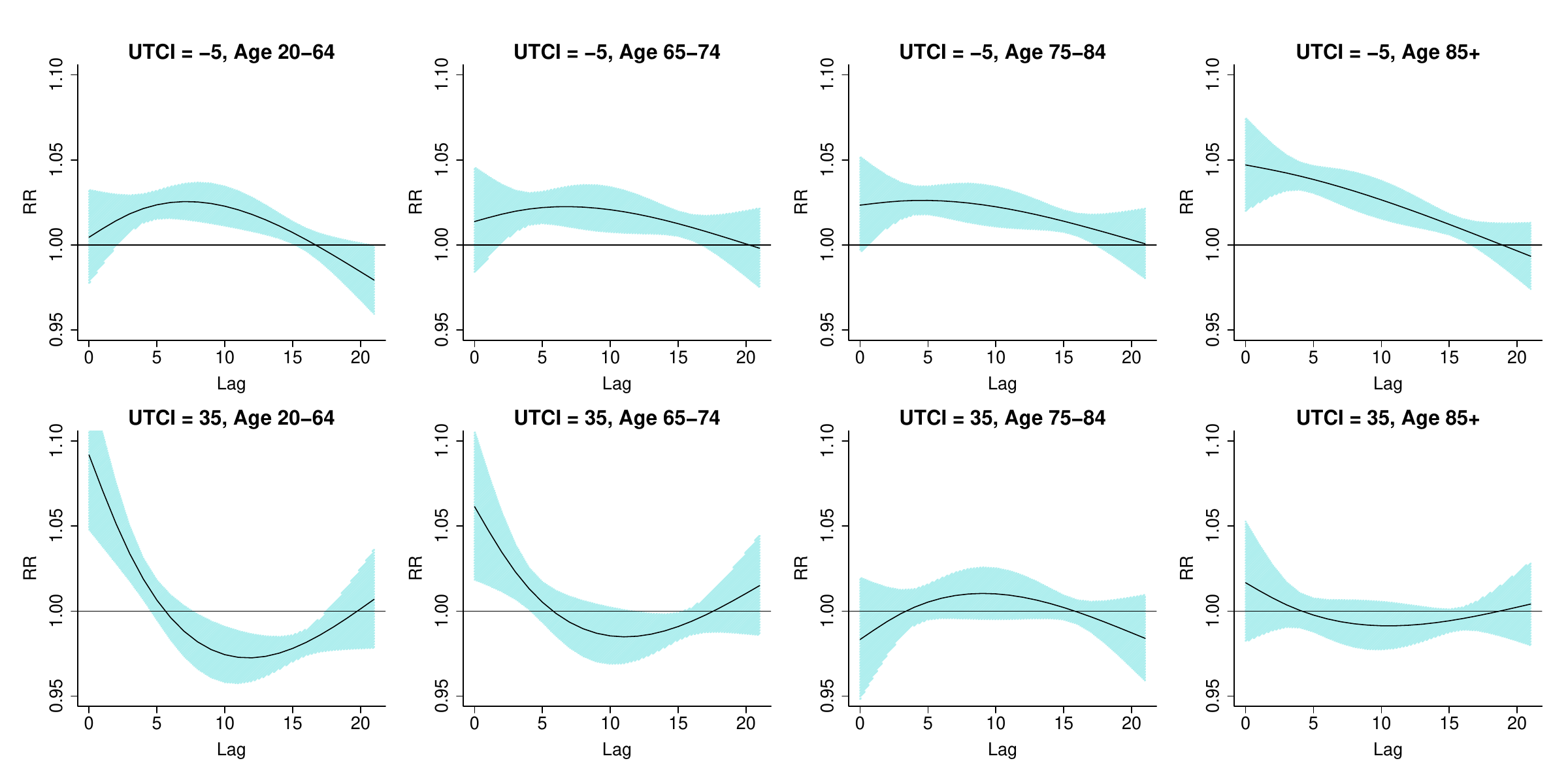}
  \caption{Lagged effect of UTCI = $-5$ and 35 in DLNM--LL model for Rome.}
  \label{fig:lag_rome_ll}
\end{figure}

\newpage
\section{Model specifications and diagnostics}
\subsection{Optimal time series model for time-varying factors}
We apply the SARIMA model for $K(t)$ and $\kappa(t,i)$ for both DLNM-LC model and DLNM-LL model. The optimal time series model is selected by \texttt{auto.arima} in \texttt{R} package \texttt{forecast}. A SARIMA model for $y_t$ with orders $(p,d,q)\times(P,D,Q)_s$ and a seasonal lag of $s$ can be expressed as:
    $$\phi_p(B)\Phi_P(B^s)(1-B)^d\left(1-B^s\right)^D y_t = \theta_q(B)\Theta_Q(B^s)\epsilon_t,$$
    where $B$ is the backward shift operator such that $B(y_t) = y_{t-1}$.   The parameters $p, d, q$ and $P, D, Q$ denote the orders of the autoregressive (AR), moving average (MA), and differencing components for the non-seasonal and seasonal parts, respectively. $s$ denotes the seasonal lag. $\{\phi_j\}_{j=1}^{p}$ and $\{\Phi_J\}_{J=1}^{P}$ denote the non-seasonal and seasonal AR coefficients, respectively, while $\{\theta_j\}_{j=1}^{q}$ and $\{\Theta_J\}_{J=1}^{Q}$ denote the corresponding MA coefficients. 

    Table \ref{tab:optimal_sarima} shows the orders of the selected optimal SARIMA models. 
    \begin{table}[H]
        \centering
        \scalebox{0.9}{
        \begin{tabular}{ccc}
        \toprule
        \small\textbf{Time series model} & \small\textbf{DLNM--LC} & \small\textbf{DLNM--LL} \\
        \midrule
        $K(t)$ & -- & ARIMA$(1,0,2)\times(0,0,1)_{52}$\\
        $\kappa(t, \text{Athens})$ & ARIMA$(1,0,0)\times(0,0,0)_{52}$ & ARIMA$(2,0,2)\times(0,0,1)_{52}$\\
        $\kappa(t, \text{Lisbon})$ & ARIMA$(1,0,0)\times(1,0,0)_{52}$ & ARIMA$(1,0,0)\times(0,0,0)_{52}$\\
        $\kappa(t, \text{Rome})$ & ARIMA$(2,0,1)\times(1,0,0)_{52}$ & ARIMA$(2,0,2)\times(1,0,0)_{52}$\\
        \bottomrule
        \end{tabular}}
        \caption{The selected optimal SARIMA model for time-varying factors.}
        \label{tab:optimal_sarima}
    \end{table}

 The estimated coefficients of the optimal SARIMA models are provided in Table \ref{tab:optimal_sarima_coef}.
    \begin{table}[H]
      \centering
      \scalebox{0.9}{
      \begin{tabular}{ccccccc}
        \toprule
        \small\textbf{DLNM--LC} & $\hat{\phi}_1$ & $\hat{\phi}_2$ & $\hat{\theta}_1$ & $\hat{\theta}_2$ & $\hat{\Phi}_1$ & $\hat{\Theta}_1$\\
        \midrule
        $\kappa(t, \text{Athens})$ & $0.282^{***}$ & $-$ & $-$ & $-$ & $-$ & $-$\\
        $\kappa(t, \text{Lisbon})$ & $0.487^{***}$ & $-$ & $-$ & $-$ & $0.150^{**}$ & $-$\\
        $\kappa(t, \text{Rome})$ & $1.067^{***}$ & $-0.103$ & $-0.842^{***}$ & $-$ & $0.102$ & $-$\\
        \midrule
        \small\textbf{DLNM--LL} & $\hat{\phi}_1$ & $\hat{\phi}_2$ & $\hat{\theta}_1$ & $\hat{\theta}_2$ & $\hat{\Phi}_1$ & $\hat{\Theta}_1$\\
        \midrule
        $K(t)$ & $0.762^{***}$ & $-$ & $-0.320^{**}$ & $0.008$ & $-$ & $0.108^{*}$\\
        $\kappa(t, \text{Athens})$ & $0.414$ & $0.247$ & $-0.055$ & $-0.299^{*}$ & $-$ & $0.145^{**}$\\
        $\kappa(t, \text{Lisbon})$ & $0.339^{***}$ & $-$ & $-$ & $-$ & $-$ & $-$\\
        $\kappa(t, \text{Rome})$ & $1.539^{***}$ & $-0.560$ & $-1.389^{***}$ & $0.435$ & $-0.066$ & $-$\\
        \bottomrule
      \end{tabular}}
      \vspace{6pt}\\
      \footnotesize{Level of significance: $0.01$:***,\quad $0.05$:**,\quad $0.1$:*}
      \caption{The coefficients of selected optimal SARIMA models on time-varying factors.}
      \label{tab:optimal_sarima_coef}
    \end{table}

\subsection{Seasonality test on model residuals}
We tested seasonality in the model residual\footnote{In this work, the terms ``residual'' and ``model residual'' mentioned refer to the difference between the fitted mortality rates from the model and the observed mortality rates.} using the Quasi-Seasonality (QS) test \citep{mann1945nonparametric, ollech2020random}, the Friedman test \citep{friedman1937use}, and the Kruskall--Wallis test \citep{kruskal1952use}.
The null hypothesis for three tests is the time series is non-seasonal. The test $p$-values for DLNM--LC and DLNM--LL are shown in Table \ref{tab:seasonal_test_DLNM_LC} and \ref{tab:seasonal_test_DLNM_LL}:
\begin{table}[H]
  \centering
  \begin{tabular}{c *{3}{ccc}}
    \toprule
    \multirow{2}{*}{\small\textbf{DLNM--LC}} 
      & \multicolumn{3}{c}{\small\textbf{Athens}}
      & \multicolumn{3}{c}{\small\textbf{Lisbon}}
      & \multicolumn{3}{c}{\small\textbf{Rome}} \\
    \cmidrule(lr){2-4}\cmidrule(lr){5-7}\cmidrule(lr){8-10}
      & \small\textbf{QS} & \small\textbf{Friedman} & \small\textbf{K--W} & \small\textbf{QS} & \small\textbf{Friedman} & \small\textbf{K--W} & \small\textbf{QS} & \small\textbf{Friedman} & \small\textbf{K--W} \\
    \midrule
    20--64 & $1.00$ & $0.36$ & $0.34$ & $1.00$ & $0.95$ & $0.81$ & $1.00$ & $0.66$ & $0.20$ \\
    65--74 & $1.00$ & $0.08$ & $0.03$ & $0.26$ & $0.53$ & $0.10$ & $0.21$ & $0.03$ & $0.01$ \\
    75--84 & $1.00$ & $1.00$ & $0.97$ & $1.00$ & $0.97$ & $0.95$ & $1.00$ & $0.93$ & $0.80$ \\
    85+    & $1.00$ & $0.48$ & $0.66$ & $1.00$ & $0.06$ & $0.04$ & $1.00$ & $0.80$ & $0.64$ \\
    \bottomrule
  \end{tabular}
  \caption{The seasonality test of model residual in DLNM--LC model}
  \label{tab:seasonal_test_DLNM_LC}
\end{table}
\begin{table}[H]
  \centering
  \begin{tabular}{c *{3}{ccc}}
    \toprule
    \multirow{2}{*}{\small\textbf{DLNM--LL}} 
      & \multicolumn{3}{c}{\small\textbf{Athens}}
      & \multicolumn{3}{c}{\small\textbf{Lisbon}}
      & \multicolumn{3}{c}{\small\textbf{Rome}} \\
    \cmidrule(lr){2-4}\cmidrule(lr){5-7}\cmidrule(lr){8-10}
      & \small\textbf{QS} & \small\textbf{Friedman} & \small\textbf{K--W} & \small\textbf{QS} & \small\textbf{Friedman} & \small\textbf{K--W} & \small\textbf{QS} & \small\textbf{Friedman} & \small\textbf{K--W} \\
    \midrule
    20--64 & $0.30$ & $0.28$ & $0.24$ & $1.00$ & $0.91$ & $0.82$ & $1.00$ & $0.58$ & $0.22$ \\
    65--74 & $1.00$ & $0.11$ & $0.03$ & $0.27$ & $0.49$ & $0.09$ & $0.30$ & $0.06$ & $0.02$ \\
    75--84 & $1.00$ & $1.00$ & $0.98$ & $1.00$ & $0.98$ & $0.96$ & $1.00$ & $0.89$ & $0.73$ \\
    85+    & $1.00$ & $0.47$ & $0.56$ & $1.00$ & $0.08$ & $0.04$ & $1.00$ & $0.75$ & $0.51$ \\
    \bottomrule
  \end{tabular}
  \caption{The seasonality test of model residual in DLNM--LL model}
  \label{tab:seasonal_test_DLNM_LL}
\end{table}

The QS and Friedman tests fail to detect residual seasonality for either DLNM–LC or DLNM–LL across regions and age groups at 5\% significance level. The Kruskal–Wallis test yields two isolated rejections (age 65–74 in Athens and Rome) for both models. As these are not corroborated by the QS or Friedman test results, we still conclude that the residuals are non-seasonal.

We clarify that the SARIMA model is not fitted to model residuals and is not used to remove seasonality from the mortality rate. Instead, the SARIMA model is applied only at the forecasting stage, where it is used on the estimated time-varying factors (\textit{e.g.}, the period indices $K(t)$ or $\kappa(t,i)$) to generate their future trajectories. 
\newpage
\section{Forecast performance under alternative heat and cold wave definitions}
To conduct a comprehensive comparison, we present out-of-sample forecasting results from the DLNM--LC and DLNM--LL models, both with and without heat and cold wave variables, and using percentile-based definitions of these variables. We modify Equations (23) and (24) as follows:
$$\text{HWD}_t = \sum_{\tau = 1 + 7(t - 1)}^{7t}\mathbbm{1}\{U_{\tau}^{\max} > Q_{1-x}(U^{\max}_{\tau})\} \times\mathbbm{1}\{U^{\max}_{\tau-1} > Q_{1-x}(U^{\max}_{\tau})\} \times\mathbbm{1}\{U^{\max}_{\tau-2} > Q_{1-x}(U^{\max}_{\tau})\},$$
$$\text{CWD}_t = \sum_{\tau = 1 + 7(t - 1)}^{7t} \mathbbm{1}\{U_{\tau}^{\min} < Q_{x}(U^{\min}_{\tau})\} \times\mathbbm{1}\{U^{\min}_{\tau-1} < Q_{x}(U^{\min}_{\tau}) \} \times\mathbbm{1}\{U^{\min}_{\tau-2} < Q_{x}(U^{\min}_{\tau}) \},$$
where $Q_x(U_\tau)$ denotes the $x\%$ percentile of $U^{\min}_\tau$ and $Q_{1-x}(U_\tau)$ denotes the $(1-x)\%$ percentile of $U^{\max}_\tau$.
In this study, we consider 2.5th, 5th, and 10th percentile. 
These results are illustrated in Tables \ref{tab:dlnm_lc_percentile} and \ref{tab:dlnm_ll_percentile}. For completeness, we also include the original LC and LL models as benchmark models.

By comparing DLNM--LC and DLNM--LL with their counterparts without the heat and cold wave variables (\textit{i.e.}, DLNM--LC (Null) and DLNM--LL (Null)), we conclude that including $\text{HWD}$ and $\text{CWD}$ in the proposed models can overall improve forecasting accuracy. Additionally, we find comparable results across all models with HWD and CWD variables under different definitions, indicating that changing the definition of these variables has no impact on the main conclusions of the study. Based on these findings, we have decided to retain the heat and cold wave variables in our proposed DLNM--LC and DLNM--LL models using the original definitions. 

\begin{table}[H]
    \centering
    \scalebox{0.9}{
    \begin{tabular}{ccccccc}
    \toprule
    \shortstack{\small\textbf{Athens}\\\textbf{}} & \shortstack{\small\textbf{LC}\\\textbf{}} & \shortstack{\small\textbf{DLNM--LC}\\\small\textbf{(Null)}} & \shortstack{\small\textbf{DLNM--LC}\\\textbf{}} & \shortstack{\small\textbf{DLNM--LC}\\\small\textbf{(2.5\%)}} & \shortstack{\small\textbf{DLNM--LC}\\\small\textbf{(5\%)}} & \shortstack{\small\textbf{DLNM--LC}\\\small\textbf{(10\%)}} \\
    \midrule
    20--64 & 0.0258 & 0.0237 & \textbf{0.0231} & 0.0241 & 0.0239 & 0.0241 \\
    65--74 & 0.1636 & 0.1352 & 0.1355 & 0.1361 & 0.1366 & \textbf{0.1341} \\
    75--84 & 0.5009 & 0.3906 & 0.3675 & 0.3860 & 0.3869 & \textbf{0.3648} \\
    85+ & 1.8120 & 1.3086 & 1.2836 & 1.3062 & 1.3283 & \textbf{1.2597} \\
    \midrule
    \shortstack{\small\textbf{Lisbon}\\\textbf{}} & \shortstack{\small\textbf{LC}\\\textbf{}} & \shortstack{\small\textbf{DLNM--LC}\\\small\textbf{(Null)}} & \shortstack{\small\textbf{DLNM--LC}\\\textbf{}} & \shortstack{\small\textbf{DLNM--LC}\\\small\textbf{(2.5\%)}} & \shortstack{\small\textbf{DLNM--LC}\\\small\textbf{(5\%)}} & \shortstack{\small\textbf{DLNM--LC}\\\small\textbf{(10\%)}} \\
    \midrule
    20--64 & 0.0274 & 0.0260 & \textbf{0.0257} & 0.0262 & 0.0268 & 0.0274 \\
    65--74 & \textbf{0.1500} & 0.1599 & 0.1592 & 0.1603 & 0.1582 & 0.1584 \\
    75--84 & \textbf{0.4615} & 0.5251 & 0.5032 & 0.5215 & 0.5200 & 0.5330 \\
    85+ & 1.8087 & 1.7477 & \textbf{1.7293} & 1.7744 & 1.7595 & 1.7713 \\
    \midrule
    \shortstack{\small\textbf{Rome}\\\textbf{}} & \shortstack{\small\textbf{LC}\\\textbf{}} & \shortstack{\small\textbf{DLNM--LC}\\\small\textbf{(Null)}} & \shortstack{\small\textbf{DLNM--LC}\\\textbf{}} & \shortstack{\small\textbf{DLNM--LC}\\\small\textbf{(2.5\%)}} & \shortstack{\small\textbf{DLNM--LC}\\\small\textbf{(5\%)}} & \shortstack{\small\textbf{DLNM--LC}\\\small\textbf{(10\%)}} \\
    \midrule
    20--64 & 0.0197 & 0.0198 & 0.0204 & 0.0195 & 0.0197 & \textbf{0.0193} \\
    65--74 & \textbf{0.1189} & 0.1412 & 0.1491 & 0.1355 & 0.1394 & 0.1403 \\
    75--84 & \textbf{0.3455} & 0.3768 & 0.3511 & 0.3611 & 0.3642 & 0.3754 \\
    85+ & 1.4920 & 1.4910 & 1.4388 & \textbf{1.3391} & 1.3795 & 1.4463 \\
    \bottomrule
    \end{tabular}
    }
    \caption{\small MAE from 10-fold expanding-window cross-validation for single-population models.}
    \label{tab:dlnm_lc_percentile}
\end{table}

\begin{table}[H]
    \centering
    \scalebox{0.9}{
    \begin{tabular}{ccccccc}
    \toprule
    \shortstack{\small\textbf{Athens}\\\textbf{}} & \shortstack{\small\textbf{LL}\\\textbf{}} & \shortstack{\small\textbf{DLNM--LL}\\\small\textbf{(Null)}} & \shortstack{\small\textbf{DLNM--LL}\\\textbf{}} & \shortstack{\small\textbf{DLNM--LL}\\\small\textbf{(2.5\%)}} & \shortstack{\small\textbf{DLNM--LL}\\\small\textbf{(5\%)}} & \shortstack{\small\textbf{DLNM--LL}\\\small\textbf{(10\%)}} \\
    \midrule
    20--64 & 0.0285 & 0.0246 & 0.0233 & 0.0243 & 0.0238 & \textbf{0.0232} \\
    65--74 & 0.1830 & 0.1606 & 0.1565 & 0.1562 & 0.1608 & \textbf{0.1494} \\
    75--84 & 0.5091 & 0.4101 & 0.3767 & 0.3801 & 0.3949 & \textbf{0.3594} \\
    85+ & 2.0056 & 1.4937 & 1.5514 & 1.3471 & 1.4345 & \textbf{1.3097} \\
    \midrule
    \shortstack{\small\textbf{Lisbon}\\\textbf{}} & \shortstack{\small\textbf{LL}\\\textbf{}} & \shortstack{\small\textbf{DLNM--LL}\\\small\textbf{(Null)}} & \shortstack{\small\textbf{DLNM--LL}\\\textbf{}} & \shortstack{\small\textbf{DLNM--LL}\\\small\textbf{(2.5\%)}} & \shortstack{\small\textbf{DLNM--LL}\\\small\textbf{(5\%)}} & \shortstack{\small\textbf{DLNM--LL}\\\small\textbf{(10\%)}} \\
    \midrule
    20--64 & 0.0274 & 0.0261 & \textbf{0.0257} & 0.0261 & 0.0266 & 0.0269 \\
    65--74 & 0.1624 & 0.1405 & \textbf{0.1382} & 0.1424 & 0.1411 & 0.1402 \\
    75--84 & 0.5012 & 0.4197 & \textbf{0.4055} & 0.4189 & 0.4065 & 0.4169 \\
    85+ & 2.0249 & 1.7761 & \textbf{1.6425} & 1.7330 & 1.6536 & 1.7393 \\
    \midrule
    \shortstack{\small\textbf{Rome}\\\textbf{}} & \shortstack{\small\textbf{LL}\\\textbf{}} & \shortstack{\small\textbf{DLNM--LL}\\\small\textbf{(Null)}} & \shortstack{\small\textbf{DLNM--LL}\\\textbf{}} & \shortstack{\small\textbf{DLNM--LL}\\\small\textbf{(2.5\%)}} & \shortstack{\small\textbf{DLNM--LL}\\\small\textbf{(5\%)}} & \shortstack{\small\textbf{DLNM--LL}\\\small\textbf{(10\%)}} \\
    \midrule
    20--64 & 0.0192 & 0.0191 & 0.0206 & \textbf{0.0185} & 0.0188 & 0.0185 \\
    65--74 & 0.1276 & 0.1251 & 0.1475 & \textbf{0.1231} & 0.1254 & 0.1272 \\
    75--84 & 0.3927 & 0.3444 & \textbf{0.3254} & 0.3327 & 0.3410 & 0.3452 \\
    85+ & 1.7384 & 1.4540 & 1.3179 & \textbf{1.2673} & 1.2859 & 1.3409 \\
    \bottomrule
    \end{tabular}
    }
    \caption{\small MAE from 10-fold expanding-window cross-validation for multi-population models.}
    \label{tab:dlnm_ll_percentile}
\end{table}
        
\clearpage
\newpage
\section{Annual mortality projections}
\begin{figure}[H]
\centering
\includegraphics[width=0.9\linewidth]{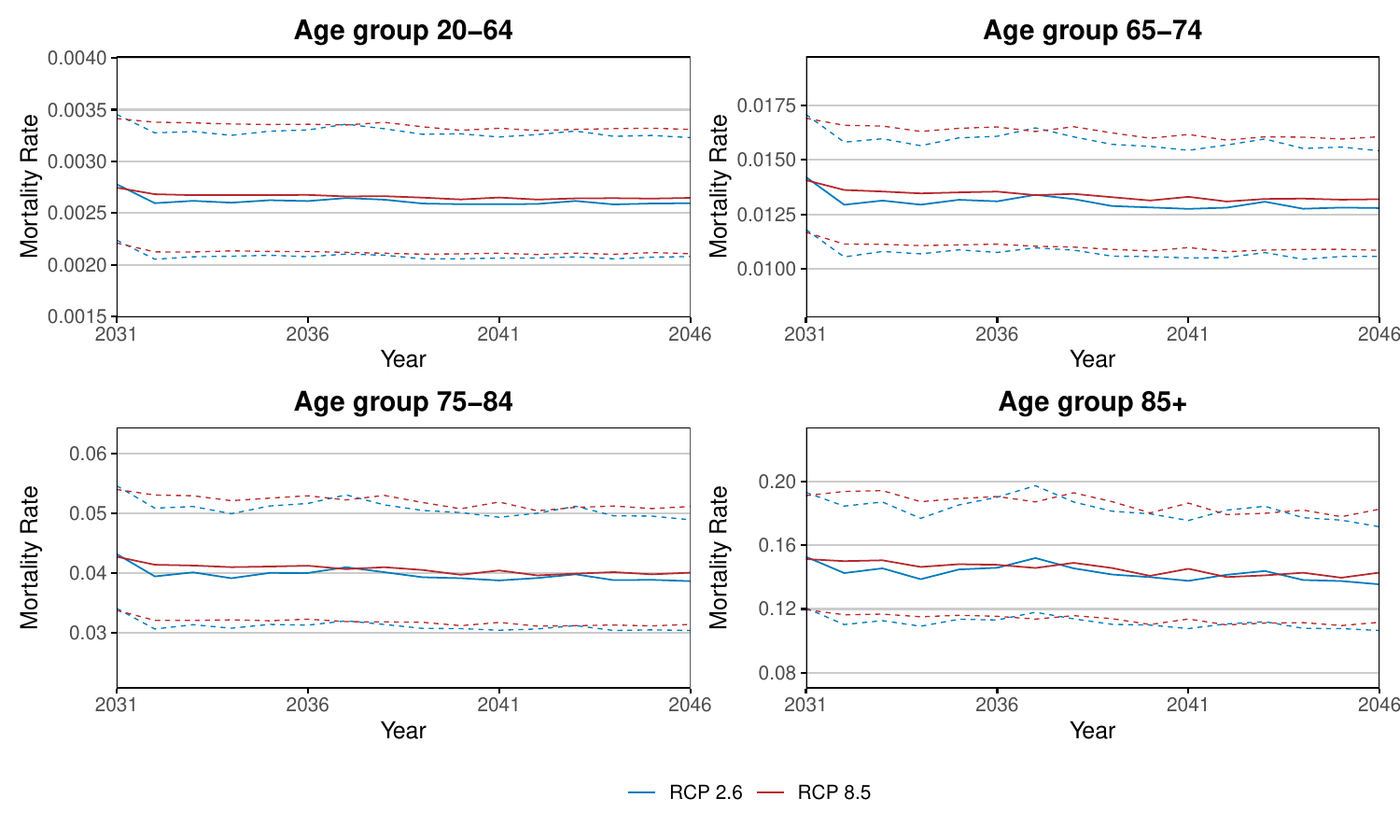}
\caption{Annualized mortality projection for Lisbon (2031--2045).}
\label{sim.year.lisbon}
\end{figure}

\begin{figure}[H]
\vspace{-0.2in}
\centering
\includegraphics[width=0.9\linewidth]{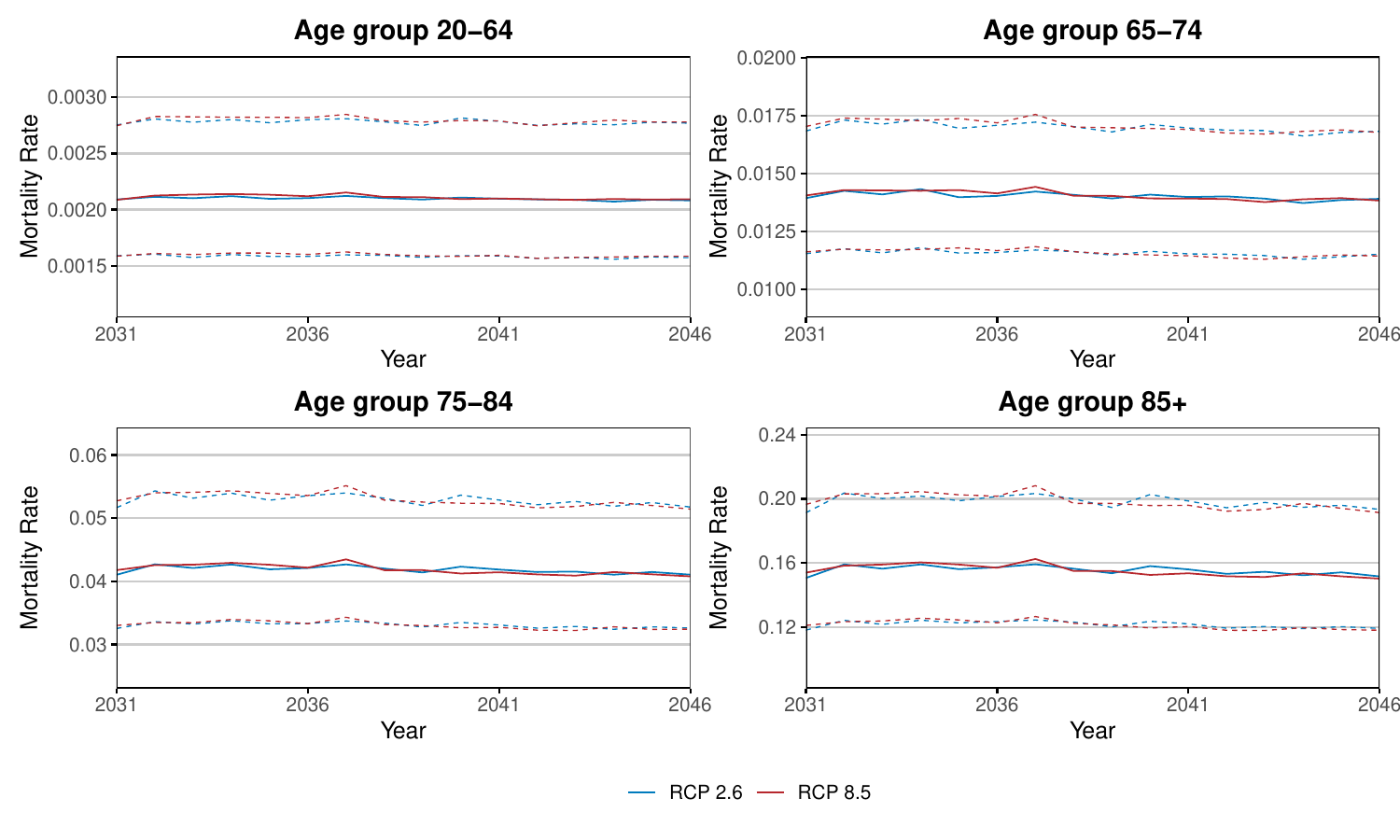}
\caption{Annualized mortality projection for Rome (2031--2045).}
\label{sim.year.rome}
\end{figure}

\newpage
\section{Comparison with benchmark models}

\subsection[Madaniyazi et al. (2024) model]{\texorpdfstring{\citeauthor{madaniyazi2024seasonality} (\citeyear{madaniyazi2024seasonality})}{Madaniyazi et al. (2024)} model}
\cite{madaniyazi2024seasonality} analyzes the relationship between daily mortality rates and temperature by combining long-term trends, baseline seasonality, and non-linear and delayed temperature effects within a DLNM-based framework. 

Since our mortality data are observed at a weekly rather than daily frequency, we retain the modeling framework of \cite{madaniyazi2024seasonality} while adapting it to the weekly setting. Specifically, for a given age $x$ and region $i$, a quasi-Poisson DLNM is fitted to the weekly death counts using a log link function as follows:
\begin{align*}
    \log \mathbb{E}[D(x,t,i)] = \log(E(x,t,i)) + \beta_0 + \beta_1 t + \text{cs}(woy(t)) + \sum_{\ell = 0}^L s(U_{\tau_t - \ell}, \ell; \bm{\nu}, \bm{\eta}),
\end{align*}
where $\log(E(x,t,i))$ is the offset term, $t$ is the weekly time index, and $\text{cs}(woy(t))$ is a cyclic spline of  week-of-year ($woy$) for week $t$, with 4 degree of freedom to capture baseline seasonality. We note that the mixed-frequency cross-basis function $\sum_{\ell = 0}^L s(U_{\tau_t - \ell}, \ell; \bm{\nu}, \bm{\eta})$ coincides with that specified in our proposed model in Section 2.1 of the manuscript. We include a linear time trend ($\beta_1 t$) to capture long-term mortality trends, analogous to the natural cubic spline with 2 degrees of freedom per decade used in \cite{madaniyazi2024seasonality}.

Based on this adaptation, we compare the out-of-sample forecasting performance of our proposed models with that of the \cite{madaniyazi2024seasonality} model using 10-fold expanding-window cross-validation, as reported in Section 4 of the manuscript.
It should be noted that \cite{madaniyazi2024seasonality} is a strong benchmark relative to the other models considered. The results thus provide reassurance that our proposed models perform well in short-term forecasting while offering a joint modeling framework across age and regions, and enabling the separation of common time trends from region-specific trends.

\subsection[Guibert et al. (2025) model]{\texorpdfstring{\citeauthor{GUIBERT2025} (\citeyear{GUIBERT2025})}{Guibert et al. (2025)} model}
We would like to reiterate that the one-step backfitting method enables us to jointly investigate the impact of climate variables and long-term mortality trends in an integrated manner. 
It should be noted that we are the first to adapt the backfitting method, which was originally developed for applications such as generalized additive models, to stochastic mortality modeling. This is a novel methodological contribution to the field, as it effectively addresses an important challenge: separating climate-driven mortality component from the general downward mortality trend (referred to in the paper as the ``inherent'' or stochastic mortality component). 

On the other hand, \cite{GUIBERT2025} employ a two-stage procedure that ignores interactions between the two components. In the first stage, the temperature-related effect is estimated from observed death counts, replacing the stochastic mortality component by a coarse deterministic trend. In the second stage, the residual deaths not explained by temperature, calculated as total observed deaths minus temperature-related deaths,  are modeled
using traditional stochastic mortality models. This procedure disregards information from the stochastic mortality dynamics when estimating the temperature component, and the inconsistent treatment of population dynamics across
stages precludes a coherent decomposition of effects.

In addition, there are other differences between \citeauthor{GUIBERT2025} model and our proposed models. To systematically highlight them, we focus on three key aspects: 
\begin{itemize}
    \item How seasonality and trends in mortality are handled;
    \item Data frequency and  availability;
    \item Forecasting accuracy and insights.
\end{itemize} 
The following subsections provide a point-by-point comparison with \cite{GUIBERT2025}.

\subsubsection{How seasonality and trends in mortality are handled}
In our method, mortality patterns are decomposed into three components: climate-driven seasonal mortality, non climate-driven seasonal mortality, and long-term mortality trend. Climate-driven seasonal mortality is captured using the DLNM model, while the other two components are captured by the stochastic mortality model. While \cite{GUIBERT2025} use a DLNM model to identify temperature-related mortality in the first stage of their modeling, seasonality of mortality is not explicitly shown or discussed in their framework. They simply adjust total annual mortality by subtracting the temperature-related mortality, which was extrapolated in the first step. In doing this, \cite{GUIBERT2025} only captures the annual mortality trend; the temperature–mortality fluctuations in winter and summer are not modeled and are largely ignored in their final results.

The two-stage procedure by \cite{GUIBERT2025} also creates an inconsistency in how non climate-driven mortality and long-term mortality trend are identified and modeled. In the first stage, these two components are modeled using a natural cubic B-spline function (see Equation 15 in \cite{GUIBERT2025}). In the second stage, stochastic annual mortality trends are fitted across age groups and regions. However, the information obtained in the first stage is not shared with or incorporated into the second stage of modeling. In contrast, our backfitting estimation avoids such inconsistencies and simultaneously handles the different components of the mortality data at the same weekly temporal resolution.

To conclude, our proposed models make a clear contribution to the literature by capturing a finer temporal resolution for both seasonality and trend, compared with the \citeauthor{GUIBERT2025} model.

\subsubsection{Data frequency and availability}
Another key difference between our research and \cite{GUIBERT2025} lies in the frequency and availability of the mortality and climate data used. It should be noted that, the first stage of \cite{GUIBERT2025} requires daily mortality data for 1980--2019 which are not publicly available, while in the second stage they switch to publicly available annual mortality data. This limits reproducibility and poses challenges for extending the analysis to other regions. The use of two different mortality data sources across stages may also introduce inconsistencies in the estimation.

In contrast, all data in our model framework are publicly available and drawn from a single source, making the models readily reproducible. We model mortality and UTCI data using our mixed-frequency DLNM framework, which integrates weekly mortality with daily UTCI without any temporal aggregation, retaining the weekly frequency in the estimation while utilizing all available information on climate. This is a novel adaptation to address the mixed-frequency problem. We further apply a backfitting algorithm to combine the mixed-frequency DLNM and stochastic mortality models in a one-step procedure on a weekly basis, enabling us to investigate weekly seasonal variations in mortality forecasts.
Compared with the \citeauthor{GUIBERT2025} model, our proposed models focus on high-frequency forecasting and can produce both weekly and annualized mortality forecasts, providing greater flexibility and more detailed insights.

\subsubsection{Forecasting accuracy and insights}
Since we do not have access to daily mortality data for the three regions considered in our empirical study, we instead adopt our proposed mixed-frequency DLNM proposed in the first stage of two-step DLNM approach. To make the forecasts from two approach comparable (\textit{i.e.} to obtain weekly mortality forecasts from \cite{GUIBERT2025}), we utilize weekly mortality data in the Li--Lee components without any data aggregation. 
Following the same estimation method in \cite{GUIBERT2025}, the estimation procedure is described as follows:
In their first stage, we apply the mixed-frequency DLNM on weekly mortality data with daily UTCI:
$$\log \mathbb{E}[y(t)] = \beta_0 + \sum_{\ell = 0}^L s(X_{\tau_t-\ell}, \ell; \bm{\nu}, \bm{\eta}) + \text{ns}(t, n_3),$$
where $\text{ns}(.)$ denotes a natural cubic B-spline.  We extract the fitted cross-basis matrix and estimate the attributable fraction $\widehat{\text{AF}}_{x,t}$ and temperature adjustment $\widehat{T}_{x,t}$, which are defined in Section 4.3 in \cite{GUIBERT2025}:
$$\widehat{T}_{x,t} = \left(1 - \widehat{\text{AF}}_{x,t}\right)^{-1} = \exp\left(\sum_{\ell = 0}^L s(X_{\tau_t-\ell}, \ell; \bm{\nu}, \bm{\eta}) \right).$$
We then estimate mortality rates that are not attributable to climate effects:
$$\widehat{\widetilde{m}}_{x,t} = \frac{m_{x,t}}{\widehat{T}_{x,t}},$$
where we denote $\widehat{\widetilde{m}}_{x,t}$ as weekly mortality rates not attributable to UTCI effects. 

In their second stage, we fit the Li--Lee model on $\widehat{\tilde{m}}_{x,t}$ at the weekly basis:
$$\log \left(\widehat{\widetilde{m}}_{x,t} \right) = A_x + B_xK_t + b_x \kappa_t + \epsilon_{x,t}.$$
Since we fit the Li--Lee model on weekly mortality rates, the seasonality of time-varying factors are considered. Thus, we apply  seasonal ARIMA model on $K_t$ and $\kappa_t$ instead of using random walk with drift or AR(1) model.

Based on the aforementioned adaptation of \citeauthor{GUIBERT2025}'s approach to our empirical study, we compare the out-of-sample forecast performance of our models to the two-step DLNM approach by \cite{GUIBERT2025}. Results of the 10-fold expanding-window cross-validation are presented in Section 4 of the manuscript. It can be seen that our proposed models outperform the \citeauthor{GUIBERT2025} model in most cases, across age groups and regions. This further confirms the advantages and improvements achieved by our approach over \cite{GUIBERT2025}.

\end{document}